
%
\magnification=1200
\vsize=24truecm
 \nopagenumbers
 \footline={ \ifnum\pageno = 1 \primapag\else\altrepag\fi}
      \def\primapag{\hss\ \hss}
       \def\altrepag{\hss\folio\hss}
\def\IC#1{\null\vskip-1.5truecm
{\rightline{IC/92/#1}\bigskip\bigskip
\centerline{International Atomic Energy Agency}
\smallskip\centerline{and}\smallskip
\centerline{United Nations Educational Scientific
and Cultural Organization}\medskip\centerline{INTERNATIONAL CENTRE FOR
THEORETICAL
PHYSICS}\vskip 2truecm  }}

\def\TITLE#1{{\centerline{\bf #1}}
}

\def\MIRAMARE{\par\vfill {\centerline{MIRAMARE -- TRIESTE}}\medskip
             \centerline{\today}\vfill \eject}

\def\AUTHOR#1{{\centerline {#1}}\smallskip}

\def\ABSTRACT{{\centerline{ABSTRACT}}\bigskip}

\parindent=40pt

\def\FOOTNOTE#1#2{\rm\parindent=0pt\footnote{#1}{#2}\parindent=40pt}
  \baselineskip=14pt
  \parskip=7pt plus 1pt

\def\today{\ifcase\month\or
             January\or February\or March \or April\or May\or June\or
             July\or August\or September\or October\or November\or December\fi
             \ \number\year}

\global\newcount\secno \global\secno=0
\global\newcount\meqno \global\meqno=1
\global\newcount\subsecno \global\subsecno=0
\def\SECTION#1{\global\advance\secno by1\global\meqno=1\global\subsecno=0
\bigbreak   \bigskip    
\noindent \hangindent 40pt {\bf\hbox to 40pt{\the\secno.\hfil}
           #1}\par\nobreak
                 \medskip\nobreak\message{#1 , }}
\def\SUBSECTION#1{\global\advance\subsecno by1\medbreak
      \noindent\hangindent 40pt
     {\bf\hbox to 40pt{\the\secno.\the\subsecno\hfil}#1}\par\nobreak
                 \medskip\nobreak\message{#1 ,  }}




\global\newcount\ftno \global\ftno=1
\def\FOOT#1{\footnote{$^{\the\ftno}$}{#1}\ %
\global\advance\ftno by1}

\global\newcount\figno \global\figno=1
\newwrite\ffile
\def\fig#1#2{\the\figno\nfig#1{#2}}
\def\nfig#1#2{\xdef#1{\the\figno}%
\ifnum\figno=1\immediate\openout\ffile=figs.tmp\fi%
\immediate\write\ffile {\noexpand \item{Fig. \noexpand#1 :\ }\noexpand#2}%
\global\advance\figno by1}
\def\semi{;\hfil\noexpand\break}

\global\newcount\tableno \global\tableno=1
\newwrite\ffile
\def\table#1#2{\the\tableno\ntable#1{#2}}
\def\ntable#1#2{\xdef#1{\the\tableno}%
\ifnum\tableno=1\immediate\openout\ffile=table.tmp\fi%
\immediate\write\ffile {\noexpand \item{Table. \noexpand#1 :\ }\noexpand#2}%
\global\advance\tableno by1}

\global\newcount\refno \global\refno=1
\newwrite\rfile
\def\ref#1#2{$^{[\the\refno]}$\nref#1{#2}}
\def\nref#1#2{\xdef#1{$^{[\the\refno]}$}%
\ifnum\refno=1\immediate\openout\rfile=refs.tmp\fi%
\immediate\write\rfile{\noexpand\item{\noexpand#1\ }\noexpand#2.}%
\global\advance\refno by1}






\def\mc{\,\raise -2.truept\hbox{\rlap{\hbox{$\sim$}}\raise5.truept
\hbox{$<$}\ }}
\def\Mc{\,\raise -2.truept\hbox{\rlap{\hbox{$\sim$}}\raise5.truept
\hbox{$>$}\ }}
%
%

%
%

%
%
\def\MIRAMARE#1{\vfill\centerline{MIRAMARE -- TRIESTE}\medskip
         \centerline{#1}\vfill}
%
%
\def\square{\sqcap\kern-6pt\lower2.4pt\hbox{--}\ }
\def\sq{\sqcap\kern-8pt\lower2.4pt\hbox{--}\ }
%
%
\def\bbone{{\rm 1}\kern-4pt{\rm 1}}

\def\bbc{{\rm C}\kern-4pt\hbox{\vrule height6.5pt width0.8pt}\ \, }

\def\bbg{{\rm G}\kern-5pt\hbox{\vrule height6pt width 0.8pt}\ \, }

\def\bbo{{\rm O}\kern-4.8pt\hbox{\vrule height6.5pt width0.8pt}\  }

\def\bbq{{\rm Q}\kern-5pt\hbox{\vrule height6pt width 0.7pt}\  \, }

\def\bbs{{\rm S}\kern-3.5pt\hbox{\vrule height6.5pt width 0.7pt}\ }
\def\bbz{{\rm Z}\!\!\! {\rm Z}}
\def\subbbc{{\rm C}\kern-3.5pt\hbox{\vrule height4.5pt width0.4pt}\, }
%

%

%

%
%
%
\def\frac#1/#2{\leavevmode\kern.1em
\raise.5ex\hbox{\the\scriptfont0 #1}
\kern-.1em/\kern-.15em\lower.25ex\hbox{\the\scriptfont0 #2}}
%


%
\def\barh{h\kern-5pt\raise3pt\hbox{-}\ }
\def\ssbarh{h\kern-4.5pt\raise3pt\hbox{-}\,}
\def\ovssbarh{h\kern-3.5pt\hbox{-}\,}


\def\longrightharpoonup{-\kern-3pt\hbox{$\rightharpoonup$}\ }

\def\centerpar{
\let\endgraf=\par \edef\restorehsize{\hsize=14truecm}
\def\par{\endgraf \restorehsize \let\par=\endgraf}
\advance\hsize by-\parindent
\item{}}

\hfuzz=15pt
\def\un{\underline{n}}
\def\uo{\underline{o}}
\def\um{\underline{m}}
\def\uk{\underline{k}}
\def\uJ{\underline{J}}
\def\ux{\underline{x}}
\def\upar{\underline\partial}
\def\bbz{{\rm Z}\kern-7pt{\rm Z}}
\IC{294}
\vskip-0.5truecm
\TITLE{GENERALIZED SPIN SYSTEMS AND $\sigma$--MODELS}
\vskip1.2truecm
\AUTHOR{S. Randjbar--Daemi}
\centerline{International Centre for Theoretical Physics, Trieste, Italy,}
\bigskip
\AUTHOR{Abdus Salam}
\centerline{International Centre for Theoretical Physics, Trieste, Italy}
\centerline{and}
\centerline{Department of Theoretical Physics, Imperial College,
London, United Kingdom}
\medskip
\centerline{and}
\medskip
\AUTHOR{J. Strathdee}
\centerline{International Centre for Theoretical Physics, Trieste, Italy.}
\vskip0.6truecm
\ABSTRACT

A generalization of the $SU(2)$--spin systems on a lattice and their
continuum limit to an arbitrary compact group $G$ is discussed. The
continuum limits are, in general, non--relativistic $\sigma$--model
type field theories targeted on a homogeneous space $G/H$, where
$H$ contains the maximal torus of $G$. In the ferromagnetic case the
equations of motion derived from our continuum Lagrangian generalize the
Landau--Lifshitz equations with quadratic dispersion relation for small
wave vectors. In the antiferromagnetic case the dispersion law is always
linear in the long wavelength limit. The models become relativistic only
when $G/H$ is a symmetric space. Also discussed are a generalization of
the Holstein--Primakoff representation of the $SU(N)$ algebra, the
topological term and the existence of the instanton type solutions in the
continuum limit of the antiferromagnetic systems.

\MIRAMARE{September 1992}

\vfill\eject

\SECTION{INTRODUCTION}

Spin systems have been studied for many years and continue to provide
new insights into the behaviour of quantum mechanical systems with
many degrees of freedom $^{1)}$. In this paper the notion of spin is
generalized to include the representations of groups larger than the
three--dimensional rotation group $^{2)}$.

The usual spin system is a collection of $SU(2)$ spin operators
associated with the sites of a lattice and coupled to their neighbours.
The Hamiltonian is
$$H={1\over 2}\ \sum_{m,n}\ J_{mn}\ S^{(m)}\cdot
  S^{(n)}+\dots                              \eqno(1.1)$$
where lattice sites are labelled by sets of integers, $m,n\dots$, and the
spin operators satisfy the $SU(2)$ commutation rules,
$$[S^{(m)}_\alpha ,S^{(n)}_\beta ]=i\ \delta_{mn}\
  \varepsilon_{\alpha\beta\gamma}\ S^{(n)}_\gamma\ .$$
In (1.1) only quadratic terms are indicated but higher order terms
could be added. Non--isotropic versions of (1.1), in which the $SU(2)$
symmetry is broken, may also be considered.

If the lattice is one--dimensional and the spin operators are
represented by Pauli matrices then the problem can be solved exactly $^{3)}$.
Otherwise, the rigorous results are only partial and it becomes necessary to
use approximate or numerical methods $^{1)}$. Two regimes are
particularly suitable for approximation: large spins and long
wavelengths $^{4)}$. In the large spin or {\it correspondence theory limit},
the spin operators are represented by unit 3--vectors,
$$S^{(m)}_\alpha\sim s\ \phi^{(n)}_\alpha ,\qquad \phi^{(n)}
  \cdot\phi^{(n)} =1,\quad s\gg 1$$
and the system becomes classical. In the long wavelength limit the
system approaches a continuum field theory appropriate for the study
of low energy excitations. When both approximations are used in
conjunction the system is described by a non--linear $\sigma$--model
$^{5)}$,
$$\eqalign{
  S^{(m)}_\alpha &\sim s\ \phi_\alpha (x)\cr
  {\cal L} &\sim {1\over 2f}\ (\partial\phi_\alpha )^2 +\dots\cr}$$
Corrections to the {\it correspondence theory limit} can be computed in the
form of a loop expansion in which the system is represented by a
so--called ``quantum'' non--linear $\sigma$--model. It is widely
believed that the quantum non--linear $\sigma$--model provides an
accurate description of the long wavelength, low energy properties of
the spin system, even in the small spin regime $^{6)}$.

A primary goal in the investigation of spin systems is to find
indications of phase transitions and critical behaviour. Thus, at
sufficiently low temperatures, it may be asked, is there long range
order? Does one find spontaneous magnetization in the ground state?
In the classical regime this is certainly true but, when quantum effects
are taken into account the ground state may well be disordered. For
example, it is known that for the one--dimensional $(D=1)$ system there
is no long range order in the ground state. On the other hand, for $D=3$
there can be long range (antiferromagnetic) order. This order is
destroyed at some finite critical temperature. For $D=2$ the ground
state has been shown to exhibit long range antiferromagnetic order for
$s\geq 1$ but the situation for $s=1/2$ is not clear, at least in the
isotropic models. Although there cannot be any long range order for $D=2$
at finite temperature,
it is interesting to consider the behaviour of the correlation length
as the temperature  goes to zero. Using renormalization group methods
it is possible to distinguish two regimes according to the coupling
strength $^{6)}$. If the coupling exceeds a certain critical value then the
correlation length remains finite for $T\to 0$, indicating a disordered
ground state. If the coupling is less than the critical value then the
correlation length diverges exponentially, indicating long range order
in the ground state $^{7)}$.

Most of the work in this field is concerned with the $SU(2)$ spin
models since it is physically motivated. Our aim here is to extend
some of these ideas to ``spin'' systems based on an arbitrary Lie group.
Instead of associating $SU(2)$ generators with the sites of a lattice
we shall use the generators from any classical Lie algebra. We shall
not attempt to generalize any of the rigorous theorems from the $SU(2)$
literature. Rather, our purpose at this stage is to develop general
formalism for treating the long wavelength and correspondence theory
limits. Our view is that while such models may not have any immediate
physical application, they may eventually serve to caste some light
on general features of the $SU(2)$ models by placing them in a broader
context.

{}From a mathematical point of view the generalized models are interesting
in themselves. In going to groups of rank $>1$ the structure becomes
much richer. For instance, one may contemplate new varieties of long
range order, beyond the familiar ferromagnetic and antiferromagnetic.
The $SU(3)$ analogue of the $s=1/2$, $SU(2)$ system would employ the
triplets $3$ and $3^*$.
In the ground state one might expect to find an orderly arrangement
of these states on the lattice. On the other hand, the large quantum
number or correspondence limit will lead to $\sigma$--models on one
or other of the manifolds, $SU(3)/SU(2)\times U(1), SU(3)/U(1)\times
U(1)$, etc. depending on the ground state.

To generalize the well--known $SU(2)$ spin wave formalism it is
necessary first of all to generalize the Holstein--Primakoff
representation of spin matrices. The approach followed here starts from
the coherent state method used by Haldane $^{8)}$ which is easy to
generalize. In this method, the finite dimensional vector spaces on
which the spin matrices act are provided with an over--complete basis
labelled by coordinates on a coset space. Transition amplitudes in this
basis are represented by path integrals whose meaning becomes
unambiguous in the limit of large quantum numbers, and which can be
used to extract a correspondence limit Lagrangian. In principle it
would also be possible to use this Lagrangian in the sub--correspondence
regime, applying canonical quantization procedures to find corrections.
However, the underlying path integral is subject to ordering
ambiguities which make the passage to quantum theory less
straightforward. This problem is solved, at least in the examples we
have examined, by extrapolating directly from the coherent basis
expectation values of the generators to operator expressions whose
correctness is then verified by using the canonical commutation rules.
The procedure will be illustrated in Sec.6 for the case of $\bbc P^N$ where
a realization of the Holstein--Primakoff type is obtained for the
generators of $SU(N+1)$.

At the classical level, where ordering problems are suppressed, it is
possible to give an explicit expression for the Lagrangian. The dynamical
variables in terms of which it is expressed can be interpreted as
coordinates on a coset space, $G/H$, where $H$ includes the Cartan
subgroup or maximal torus of $G$. If $H$ is precisely the Cartan
subgroup then $G/H$ is a flag manifold. The detailed structure of the
Lagrangian depends on the coherent state basis from which one starts.
This basis is generated by applying finite transformations, $L_\phi$,
belonging to the group $G$, to some chosen reference state, $\vert
\Lambda >$,
$$\vert\phi >\ =L_\phi\ \vert\Lambda >,\qquad L_\phi\in G\ .$$
The stability group, $H$, is defined as the set of transformations which
leave the reference state invariant, up to a phase
$$h\vert\Lambda >\ =\vert\Lambda >\ e^{i\psi (h)},
  \qquad h\in H\subset G\ .$$
The coordinates $\phi^\mu ,\ \mu =1,2, . . ,\dim G/H$, may be chosen
in any convenient way to parametrize representatives, $L_\phi$, from
the cosets $G/H$. The Cartan components of the spin connection on
$G/H$ are defined by the 1--form
$$L^{-1}_\phi\ dL_\phi =-A^j(\phi )\ H_j+\dots$$
where the operators $H_j$ are generators of the Cartan algebra. These
are among the generators of the stability group, $H$, and their
eigenvalues serve to label the reference state,
$$H_j\ \vert\Lambda >\ =\vert\Lambda >\ \Lambda_j\ .$$

The coherent basis system is established at each site on the lattice
and their direct products define an over--complete basis for the
Hilbert space of the model. It will be shown in Sec.2 that the
correspondence limit Lagrangian takes the form
$$L={\barh\over i}\ \sum_n\ A_\mu (\phi_n)\ \partial_t\
  \phi^\mu_n-H(\phi )                                     \eqno(1.2)$$
where
$$A_\mu (\phi_n) = A^j_\mu (\phi_n)\ \Lambda_{nj}$$
and
$$\eqalign{
  H(\phi ) &=\ <\phi\vert H\vert\phi >\cr
  &={1\over 2}\ \sum_{m,n}\ J_{mn}\ Q_\alpha (\phi_m)\
  Q_\alpha (\phi_n)+\dots\cr}$$
where the functions $Q_\alpha (\phi )$ are defined as coherent state
expectation values of the generators of $G$,
$$Q_\alpha (\phi_m) =\ <\phi\vert Q^{(m)}_\alpha\vert\phi >\ .$$

Generalized spin waves are obtained by examining weak excitations of the
system described by (1.2). These are defined, at the classical level,
when the coherent state at each site on the lattice is close to its
reference value. If the Lagrangian is translation invariant then the
usual Fourier techniques can be used to extract the spectrum. Some
examples of this will be given in Sec.3 where the question of classical
stability is considered.

Still at the classical level, it is straightforward to make the restriction
to long wavelength configurations, turning the lattice model into a
continuum field theory. What emerges is a kind of non--linear
$\sigma$-model with fields targeted on the coset space, $G/H$. The
Lagrangian will be first order in the time derivative but second (and
higher) order in the space derivatives. However, if the (classical)
ground state is ``antiferromagnetic'' in the sense that the reference
state weights, $\Lambda_{nj}$, alternate in sign across the lattice,
then some of the dynamical variables are algebraic and can be eliminated
to give a Lagrangian which is second order in the time derivative. This
Lagrangian is generally non--relativistic, even when higher order
space derivatives are neglected, unless $G/H$ happens to be a symmetric
space. This will be discussed in Sec.4.

In passing from the lattice to the continuum description, a term
appears in the Lagrangian density (for the antiferromagnetic case)
which is a total derivative. Such a term has no relevance at the classical
level since it makes no contribution to the equations of motion. It is
only an artefact of the method and may be safely discarded. In the
quantized continuum theory, on the other hand, such a term may not be
negligible. If it makes a finite contribution to the action functional then
it will affect the sum over configurations. Since our discussion, in
Sec.4, of the passage to the continuum theory is couched in classical
terms -- the factor ordering question is ignored -- we cannot determine
whether such terms are actually needed for the long wavelength
description of spin systems at the quantum level. Haldane observed this
term in his treatment of $SU(2)$ spin systems in one space
dimension $(D=1)$ and he noted that it has a topological interpretation,
giving an alternating sign factor in the sum over configurations when
the spin is a half--integer $^{9),2)}$. Within the limits of our discussion
we have confirmed that this effect persists in the generalized
antiferromagnetic $D=1$ spin systems. (For $D>1$ our total derivative
term does not seem to have a topogical significance and should
therefore probably be suppressed in the quantum theory $^{10)}$.)
Topological aspects of the $D=1$ systems are discussed in Sec.5 where
first order equations for generalized instantons are derived.

The classical non--linear $\sigma$--models obtained by this approach
can be quantized in the usual way. It is not at all obvious that such
quantum non--linear $\sigma$--models have much relevance to the
original spin problem but, as mentioned above, there is a common belief
in the $SU(2)$ case that they are good for describing the long
wavelength, low energy features of the spin problem. Some support for
this view may perhaps derive from the standard treatment of
quantized $\sigma$--models by dimensional regularization. In such
treatments it is prescribed that quantum corrections associated with
factor ordering should be discarded.
Recall that it is precisely these
ordering contributions -- defined by the Holstein--Primakoff
operator valued expressions for the generators $Q_\alpha (\phi_n)$ --
which distinguish the sub--correspondence regime on the lattice. If they are
indeed not relevant in the continuum limit then the quantum non--linear
$\sigma$--model should be appropriate for the long wave behaviour
of the lattice model.

Finally, it may be remarked that the loop expansions of the quantized
theory comprise quantum corrections to the classical spin wave
amplitudes. Since both $A_\mu (\phi )$ and $Q_\alpha (\phi )$ in (1.2)
are linear in the weights, $\Lambda_j$, it is clear that the loop
expansions can be read as expansions in powers of $1/\Lambda$ if the
couplings $J_{mn}$ are redefined to absorb one power of $\Lambda$,
$J_{mn}=J'_{mn}/\Lambda$. With this interpretation, the classical
theory emerges when $\Lambda\to\infty$. This is what is meant by
the expression, large quantum number or correspondence limit.

\SECTION{THE SEMICLASSICAL REGIME}

Our purpose is to discuss the large quantum number, or correspondence limit
of generalized spin systems. Typically, the Hamiltonian is given as a sum of
terms in which operators associated with sites on a lattice are coupled,
$$H={1\over 2}\ \sum_{m,n}\ J^{\alpha\beta}_{mn}\
  Q^{(m)}_\alpha\ Q^{(n)}_\beta +\dots                      \eqno(2.1)$$
where $m,n,\dots$ label the sites and the coefficients
$J^{\alpha\beta}_{mn}$ are coupling parameters. The matrices $Q^{(m)}_a$
are generators of some representation of a Lie group, $G$. They satisfy
the commutation rules
$$[Q^{(m)}_\alpha ,Q^{(n)}_\beta ]=\delta_{mn}\ c_{\alpha\beta}\
  ^\gamma\ Q^{(n)}_\gamma\ .                                     \eqno(2.2)$$
The couplings are assumed to be $G$--invariant, i.e. proportional to the
Killing metric,
$$J^{\alpha\beta}_{mn}=J_{mn}\ g^{\alpha\beta}        \eqno(2.3)$$
and also translation invariant,
$$J_{mn}=J_{m-n}\ .
\eqno(2.4)$$
Higher order and/or non--invariant coupling terms could be included
in the Hamiltonian (2.1) but we shall not consider such possibilities.

We are not looking for exact solutions. Following Affleck $^{2)}$, our aim
is to find a Lagrangian description which can be used for
semi--classical approximations, generalized spin waves, and which
leads to various $\sigma$--model type theories in the continuum limit.
Our approach is to generalize the method used by Haldane $^{3)}$.
As outlined in Sec.1 this involves the introduction of an over--complete
basis of coherent states in the finite dimensional vector space on which
the $Q_\alpha$ act, followed by the construction of a path integral
representation for transition amplitudes in the coherent basis. In the
$SU(2)$ case the coherent states in question are associated with the
points of the manifold $SU(2)/U(1)$ and they describe the orientation
of the spin vector in the correspondence theory limit. This picture
generalizes easily to the cosets $G/H$.

Let $G$ be a Lie group and $H$ one of its subgroups. We are interested
in those subgroups which include at least the Cartan subgroup
(maximal Abelian subgroup). If $H$ is precisely the Cartan subgroup then
$G/H$ is a flag manifold, but we may consider larger, non--Abelian
subgroups so that $G/H$ becomes a subspace of the flag manifold.

Suppose that $G$ is decomposed into left cosets with respect to the
subgroup, $H$. From each coset one can choose a representive element
$L_\phi\in G$ where $\phi =\{ \phi^\mu\}$ labels a point on the
manifold $G/H$. This means that, for arbitrary $g\in G$, there is a map
$$\phi\to\phi '=\phi '(\phi ,g)$$
defined by
$$gL_\phi =L_{\phi '}\ h
\eqno(2.5)$$
where $h=h(\phi ,g)\in H$. The detailed form of the functions
$\phi '(\phi ,g)$ and $h(\phi ,g)$ depends on the choice of representative
elements, $L_\phi$.

In the vector space that carries one of the irreducible representations
of $G$, choose one vector, $\vert\Lambda >$, that is invariant, up to a
phase, under the action of $H$,
$$h\vert\Lambda >\ = \vert\Lambda >\ e^{i\psi (h)}\ .   \eqno(2.6)$$
This vector will be referred to as the ``reference state''. Now,
corresponding to the points $\phi\in G/H$, define the coherent states,
$$\vert\phi >\ = L_\phi\ \vert\Lambda >\ .                    \eqno(2.7)$$
Under the action of $G$ these states are transformed according to
$$g\vert\phi >\ = \vert\phi '>\ e^{i\psi (h)}                   \eqno(2.8)$$
where $\phi '$ and $h$ are determined by (2.5).

The coherent states can be expanded in an orthonormal basis,
$$\vert\phi >\ =\sum_\lambda\ \vert\lambda >\ <\lambda\vert
  L_\phi\vert\Lambda >
\eqno(2.9)$$
and it is possible to project the orthonormal basis vectors from the
coherent states by integrating over the coset manifold,
$$\vert\lambda >\ =\int d\mu (\phi )\ \vert\phi >\ <\Lambda
  \vert L^{-1}_\phi\ \vert\lambda >                              \eqno(2.10)$$
where $d\mu$ is a suitably normalized $G$--invariant measure on
$G/H$. The coherent states therefore constitute an over--complete
basis.

Of particular interest is the overlap between neighbouring states,
$$\eqalignno{
  <\phi +d\phi\vert\phi >\ &=\ <\Lambda\vert L^{-1}_{\phi +d\phi}\ L_\phi\
    \vert\Lambda >\cr
    &=\ \Lambda\vert (1-L^{-1}_\phi\ dL_\phi )\vert\Lambda >\cr
    &= 1+ A^j\ \Lambda_j                                         &(2.11)\cr}$$
where the 1--forms, $A^j=d\phi^\mu\ A^j_\mu (\phi )$, are the Cartan
components of the spin connection on $G/H$. To obtain this we have
expanded the Maurer--Cartan form in a basis of the Lie algebra of $G$,
$$\eqalignno{
  L^{-1}_\phi\ dL_\phi &= e^\alpha\ Q_\alpha\cr
  &=-A^j\ H_j+e^{\bar\alpha}\ Q_{\bar\alpha} +
  e^{\dot\alpha}\ Q_{\dot\alpha}                         &(2.12)\cr}$$
where the operators $Q_\alpha$ satisfy the commutation rules
$$[Q_\alpha ,Q_\beta ]=c_{\alpha\beta}\ ^{\gamma}\ Q_\gamma\ .

              \eqno(2.13)$$
The generators of the Cartan algebra are denoted, $H_j$. These operators
are all in the algebra of $H$ together (in general) with non--Abelian
elements, $Q_{\bar\alpha}$. The remaining generators are denoted by
$Q_{\dot\alpha}$. Their coefficient 1--forms, $e^{\dot\alpha}$, in the
Maurer--Cartan form define the frames on $G/H$. The requirement (2.6)
that $\vert\Lambda >$ be invariant, up to a phase, under the action of
$H$ means
$$\eqalignno{
  H_j\ \vert\Lambda >\ &= \vert\Lambda >\ \Lambda_j\cr
  Q_{\bar\alpha}\ \vert\Lambda >\ &= 0\ .                     &(2.14)\cr}$$
The operators $Q_{\dot\alpha}$ belong to some representation of $H$.
This representation is often reducible but it cannot contain any
components that are neutral with respect to the Cartan algebra.
Hence\break
$<\Lambda\vert Q_{\dot\alpha}\vert\Lambda >\ =0$ and
the result (2.11) follows.

The infinitesimal formula (2.11) can be integrated to obtain a
path integral representation for the finite overlap. To do this one
makes repeated use of the completeness condition (2.10) or
$$1=\int d\mu (\phi )\ \vert\phi ><\phi\vert\ .      \eqno(2.15)$$
Thus, one writes firstly
$$\eqalign{
  <\phi '\vert\phi >\ &= \int <\phi_N\vert\phi_{N-1}>\ d\mu (\phi_{N-1})\
    <\phi_{N-1}\vert\phi_{N-2}>\ d\mu (\phi_{N-2})\dots\cr
    &\qquad\qquad\qquad \dots d\mu (\phi_1)\ <\phi_1\vert\phi_0>\cr}$$
where $\phi_N=\phi '$ and $\phi_0=\phi$. The intermediate points,
$\phi_{N-1},\phi_{N-2},\dots ,\phi_1$ are integrated over $G/H$. In the
limit $N\to\infty$ this gives rise in the usual formal way to a path
integral over configurations $\phi (t)$. The factors in the integrand
are given by (2.11),
$$\eqalign{
  <\phi +d\phi\vert\phi >\ &= \exp (A^j\ \Lambda_j)\cr
  &=\exp (dt\ \dot\phi^\mu\ A^j_\mu (\phi )\ \Lambda_j)\cr}$$
to first order in $dt$. Hence, the finite overlap is
$$<\phi '\vert\phi >\ =\int (d\mu )\exp\int dt\ \dot\phi^\mu\
A^j_\mu (\phi )\ \Lambda_j\ .                                     \eqno(2.16)$$
Finally, the coherent state transition amplitudes are represented by
the path integrals
$$<\phi '\vert\exp \left( -{i\over\barh}\ tH\right)\vert\phi  >\ =
  \int (d\mu )\exp\left( {i\over\barh}\ \int dt\ L\right)   \eqno(2.17)$$
where the Lagrangian is given by
$$L={\barh\over i}\ \dot\phi^\mu\ A^j_\mu (\phi )\
  \Lambda_j-H(\phi )\ .
\eqno(2.18)$$
The classical Hamiltonian, $H(\phi )$, is defined by the diagonal
elements of $H$ in the coherent basis,
$$H(\phi )=\ <\phi\vert H\vert\phi >\ .
\eqno(2.19)$$
The Euler Lagrange equations deriving from (2.18) take the form
$${\barh\over i}\ \Lambda_j\ F_{\mu\nu}^j\ \dot\phi^\nu =
  {\partial H\over\partial\phi^\mu}
\eqno(2.20)$$
where $F^j_{\mu\nu}=\partial_\mu\ A^j_\nu -\partial_\nu\ A^j_\mu$
are the Cartan components of the curvature tensor on $G/H$.

The path integral representation (2.17) is of course only formal.
To give it a precise meaning one must apply canonical quantization
methods to the classical Lagrangian (2.18) and invent a suitable
prescription for interpreting the ambiguous factor ordering in
$H(\phi )$. The canonical momenta are define by
$$\eqalignno{
  \pi_\mu &= {\partial L\over\partial \dot\phi^\mu}\cr
  &= {\barh\over i}\ A^j_\mu (\phi )\ \Lambda_j\ .            &(2.21)\cr}$$
To proceed further with this it will be necessary to choose some
parametrization of $G/H$ and obtain explicit formulae for the
components $A^j_\mu$. Once the commutation rules have been
determined for the $\phi^\mu$, it becomes possible to consider the
structure of $H(\phi )$. For the systems we are concerned with, the
Hamiltonian is given as a polynomial in the generators $Q_\alpha$.
This means that the classical Hamiltonian will be expressed in terms
of the functions
$$Q_\alpha (\phi ) =\ <\phi\vert Q_\alpha\vert\phi >\ .    \eqno(2.22)$$
These functions are unambiguous at the classical level but when the
dynamical variables $\phi^\mu$ are quantized it is necessary to
determine the factor ordering such that the algebra is realized,
$$[Q_\alpha (\phi ),Q_\beta (\phi )]=c_{\alpha\beta}\ ^\gamma\
  Q_\gamma (\phi )\ .$$
This is precisely the problem solved by Holstein and Primakoff $^{11)}$
for the case of $SU(2)$.  In Sec.6 we shall give a generalization for
the case of $SU(N+1)$ with coordinates $\phi^\mu$ on $\bbc P^N$.
There it will be verified that the
$\phi$--commutators vanish like $1/\Lambda$ in the limit $\Lambda\to
\infty$ (see Eq.(6.10)). But
first we shall continue with the classical approximation (i.e. the large
quantum number limit).

\SECTION{WEAK EXCITATIONS AND CLASSICAL STABILITY}

Suppose that the Hamiltonian for the generalized spin system on a
lattice is given as a sum over pairs of sites, etc., as described in Sec.1.
In the classical approximation,
$$H(\phi )={1\over 2}\ \Sigma\ J_{mn}\ g^{\alpha\beta}\
  Q_\alpha (\phi_m)\ Q_\beta (\phi_n)+\dots                        \eqno(3.1)$$
where $g^{\alpha\beta}$ is the Killing metric and the functions
$Q_\alpha (\phi )$ are defined by (2.22). It is convenient to express
these functions in terms of matrix elements of $L_\phi$ in the adjoint
representation. These matrices, $D_\alpha\ ^\beta$, are defined by
$$g^{-1}\ Q_\alpha\ g=D_\alpha\ ^\beta (g)\ Q_\beta ,\quad g\in G\ .$$
If follows from the definition (2.7) of the coherent states that
$$\eqalignno{
  Q_\alpha (\phi ) &=\ <\Lambda\vert L^{-1}_\phi\ Q_\alpha\
  L_\phi\ \vert\Lambda >\cr
  &= D_\alpha\ ^\beta (L_\phi )\ <\Lambda\vert Q_\beta\vert
  \Lambda >\cr
  &=D_\alpha\ ^j(L_\phi )\ \Lambda_j\ .      &(3.2)\cr}$$
Hence (3.1) takes the form
$$\eqalignno{
  H(\phi ) &= {1\over 2}\ \Sigma\ J_{mn}\ g^{\alpha\beta}\
  D_\alpha\ ^j(L_{\phi_m})\ D_\beta\ ^k(L_{\phi_n})\
  \Lambda_{mj}\ \Lambda_{nk} +\dots\cr
  &= {1\over 2}\ \Sigma\ J_{mn}\ \Lambda^j_m\ D_j\ ^k
  (L_{\phi_m}^{-1}\ L_{\phi_n})\Lambda_{nk} +\dots         &(3.3)\cr}$$
where the invariance of the metric has been used,
$$\eqalign{
  g^{\beta\alpha}\ D_\alpha\ ^j(L) &= g^{j\alpha}\ D_\alpha\ ^\beta
  (L^{-1})\cr
  &=g^{jk}\ D_k\ ^\beta (L^{-1})\ .\cr}$$
The indices $j,k$ refer to components in the Cartan algebra, and we
take a basis such that the metric tensor is block diagonal.

Since the coset manifold is homogeneous, no generality is lost by
assuming that $\phi_n$ becomes independent if $n$ in the ground
state, i.e.
$$<L^{-1}_{\phi_m}\ L_{\phi_n}>\ =1\ .                             \eqno(3.4)$$
With this convention the ground state value of the Hamiltonian (3.3)
reduces to
$$<H(\phi )>\ = {1\over 2}\ \sum_{mn}\ J_{mn}\
  \Lambda_m\cdot\Lambda_n+\dots                                  \eqno(3.5)$$
Minimization of this expression must be our guide in choosing the
weights $\Lambda_n$ that define the ground state configuration.
To be more specific, we shall suppose that the pattern of weights
takes the form of a finite number of translation invariant sublattices.
Write the variables in the form
$$\{\phi_n\} =\{\phi_{\un\nu}\}
\eqno(3.6)$$
where $\un =(n_1,n_2,\dots ,n_D)$ is a $D$--vector with integer
components and $\nu =1,\dots ,f$ indicates the sublattice. The weights
are independent of $\un$,
$$\Lambda_{\un\nu} =\Lambda_\nu\ .$$
The translation invariant coupling parameters $J_{\um\mu , \un\nu}$
depend on $\um -\un$. The ground state energy per cell is then given by
$${1\over 2}\ \sum_{\un ',\nu ,\nu '}\ J_{\un\nu ,\uo\nu '}\
  \Lambda_\nu\cdot\Lambda_{\nu '}\ .                         \eqno(3.7)$$

To minimize the ground state energy density in, for example, a
one--dimensional system with nearest neighbour couplings one should
choose $\Lambda_n\cdot\Lambda_{n+1}$ to be positive (negative)
if $J_{n,n+1}$ is negative (positive). In this way one is led to the most
obvious generalizations of ferromagnetic (antiferromagnetic) order.
In more than one dimension there are new possibilities. For example,
in two dimensions with a triangular lattice each site has six nearest
neighbours. If all nearest neighbours are coupled with the same
positive strength then one can make an ordered ground state with
three distinct sublattices such that the energy density is
$$J(\Lambda_1\Lambda_2+\Lambda_2\Lambda_3+\Lambda_3
  \Lambda_1)={J\over 2}\left[ (\Lambda_1+\Lambda_2+
  \Lambda_3)^2-\Lambda^2_1-\Lambda^2_2-\Lambda^2_3\right]\ .$$
Hence the configuration with $\Lambda_1+\Lambda_2+\Lambda_3=0$
would be favoured. Such a generalized ``antiferromagnet'' becomes a
possibility in the $SU(3)$ models.

To test the stability of conjectured ground state configurations one
can compute the energy associated with weak perturbations and find
the excitation spectrum. Write
$$\phi_n =\phi +\Delta\phi_n                                       \eqno(3.8)$$
where $\phi$ is a constant and $\Delta \phi_n$ is small. Substitute
into (3.3) and collect the bilinear terms. To carry out this
computation we need the formula
$$\eqalign{
  L^{-1}_\phi\ L_{\phi +\Delta\phi} &= 1+\Delta\phi^\mu\ L^{-1}\
  \partial_\mu\ L +{1\over 2}\ \Delta\phi^\mu\ \Delta\phi^\nu\
  L^{-1}\ \partial_\mu\partial_\nu\ L +\dots\cr
  &= 1+\Delta\phi^\mu\ e_\mu\ ^\alpha\ Q_\alpha + {1\over 2}\
  \Delta\phi^\mu\ \Delta\phi^\nu (\partial_\mu\ e_\nu\ ^\alpha\
  Q_\alpha + e_\mu\ ^\alpha\ e_\nu\ ^\beta\ Q_\alpha\ Q_\beta )+\dots\cr}$$
which derives from (2.12). Hence,
$$\eqalign{
  L^{-1}_{\phi_m}\ L_{\phi_n} &= 1+\Biggl[ (\Delta\phi_n\ ^\mu -
  \Delta\phi_m\ ^\mu )\ e_\mu\ ^\alpha + {1\over 2} (\Delta\phi_n\ ^\mu\
  \Delta\phi_n\ ^\nu -\Delta\phi_m\ ^\mu\ \Delta\phi_m\ ^\nu )
  \partial_\mu\ e_\nu\ ^\alpha -\cr
  &\quad -{1\over 2}\ \Delta\phi_m\ ^\mu\ \Delta\phi_n\ ^\nu\
  e_\mu\ ^\beta\ e_\nu\ ^\gamma\ c_{\beta\gamma}\ ^\alpha
  \Biggr] Q_\alpha\cr
  &\quad +{1\over 2} (\Delta\phi_n\ ^\mu -\Delta\phi_m\ ^\mu )
  (\Delta\phi_n\ ^\nu -\Delta\phi_m\ ^\nu ) e_\mu\ ^\alpha\
  e_\nu\ ^\beta\ Q_\alpha Q_\beta +\dots\cr}$$
and, in the adjoint representation,
$$D_j\ ^k(L^{-1}_{\phi_m}\ L_{\phi_n}) =\delta^k_j +{1\over 2}
  (\Delta\phi_n\ ^\mu -\Delta\phi_m\ ^\mu )
  (\Delta\phi_n\ ^\nu  -\Delta\phi_m\ ^\mu )
  e_\mu\ ^\alpha\ e_\nu\ ^\beta\ c_{j\alpha}\ ^\gamma\
  c_{\gamma\beta}\ ^k +\dots$$
With this approximation the Hamiltonian (3.3) takes the form
$$H(\phi) ={1\over 2}\ \Sigma\ J_{mn}\left[\Lambda_m\cdot
  \Lambda_n+{1\over 2}\ (\Delta\phi_n\ ^\mu -\Delta\phi_m\ ^\mu )
  (\Delta\phi_n\ ^\nu -\Delta\phi_m\ ^\nu) k_{\mu\nu}^{mn}
  (\phi )+\dots\right]
\eqno(3.9)$$
where $k_{\mu\nu}$ is a symmetric $G$--invariant tensor on $G/H$
defined by
$$k^{mn}_{\mu\nu} =e_\mu\ ^\alpha\ e_\nu\ ^\beta\ \Lambda^j_m\
  c_{j\alpha}\ ^\gamma\ c_{\gamma\beta}\ ^k\
  \Lambda_{nk}\ .
\eqno(3.10)$$
The ground state will be stable against weak classical perturbations if
the matrix, $J_{mn}\ k^{mn}_{\mu\nu}$, is positive. To examine this
question it is helpful to express the algebra of $G$ in the Cartan--Weyl
basis with generators $H_j$ and $E_\alpha$, where $\alpha$ denotes
a root. The commutation rules include
$$\eqalign{
  [H_j, E_{\pm\alpha}] &=\pm\ \alpha_j\ E_{\pm\alpha},\cr
  [E_\alpha ,E_{-\alpha}] &= \alpha^j\ H_j\ .\cr}$$
Since we are dealing with unitary representations we can choose the
basis such that
$$H_j\ ^+=H_j\quad {\rm and}\quad E_\alpha\ ^+= E_{-\alpha}\ .$$
The root vectors are then real. However, it is important to remark that
since $L^{-1}dL$ is antihermitian, the frame components in the
Cartan--Weyl basis must satisfy
$$(\Delta\phi^\mu\ e_\mu\ ^\alpha )^* =-\Delta\phi ^\mu\
  e_\mu\ ^{-\alpha}\ .$$
The tensor (3.10) reduces in this basis to a sum over roots,
$$k^{mn}_{\mu\nu} =\sum_{roots}\ e_\mu\ ^\alpha\
  e_\nu\ ^{-\alpha}\ \Lambda_m\cdot \alpha\
  \alpha\cdot\Lambda_n\  .
\eqno(3.11)$$
Therefore,
$$\Delta\phi^\mu\ \Delta\phi^\nu\ k^{mn}_{\mu\nu} =-
  \sum_{roots}\ \vert\Delta\phi^\mu\ e_\mu\ ^\alpha\vert^2\
  \Lambda_m\cdot\alpha\ \alpha\cdot\Lambda_n\ .$$
It follows that the conjectured ground state will be stable if the
matrix, $J_{mn}\ \Lambda_{mj}\ \Lambda_{nk}$, is negative definite.
This is a sufficient condition. It would be straightforward in principle
to sharpen this criterion for a specific model but it would not be very
useful to pursue the question in generality.

We conclude this discussion with brief remarks about the excitation
spectrum. To simplify the notation define the frame components of
the weak fluctuations,
$$\Delta\phi_n\ ^\mu\ e_\mu\ ^\alpha =\psi_n^\alpha =-
  (\psi_n^{-\alpha})^*\ .
  \eqno(3.12)$$
In terms of these variables the bilinear part of the Hamiltonian
(2.31) becomes
$$H_2=-{1\over 2}\ \sum_{\alpha >0}\ \sum_{m,n}\ J_{mn}\
 \Lambda_m\cdot\alpha\ \alpha\cdot\Lambda_n\
 \vert\psi_n^\alpha -\psi_m^\alpha\vert^2                        \eqno(3.13)$$
where the sum is restricted to positive roots. With the translation
invariant background, described above, comprising a finite number of
sublattices, $\{ n\} =\{\un ,\nu\}$, it is natural to use Fourier series.
Define the Fourier components,
$$\eqalignno{
  \tilde\psi^\alpha_\nu (\uk) &=\sum_{\un}\ \psi^\alpha_{\un\nu}\
  e^{-i\uk\cdot\un}\cr
  \psi^\alpha_{\un\nu} &=\int\left({dk\over 2\pi}\right)^D\
  \tilde\psi^\alpha_\nu (\uk)\ e^{i\uk\cdot\un}                  &(3.14)\cr}$$
where the integration ranges over a cell of volume $(2\pi )^D$. The
Hamiltonian (3.13) then takes the form
$$\eqalignno{
  H_2 &=-{1\over 2}\ \sum_{\alpha >0}\int\left({dk\over 2\pi}
  \right)^D\ \sum_{\un}\ \sum_{\nu ,\nu '}\cdot\cr
  &\quad\cdot J_{\un\nu ,\uo\nu '}\ \Lambda_\nu\cdot\alpha\
  \alpha\cdot\Lambda_{\nu '}\ \big\vert\tilde\psi^\alpha_\nu (\uk )\
  e^{i\uk\cdot\un /2}-\tilde\psi^\alpha_{\nu '} (\uk )\
  e^{-i\uk\cdot\un /2}\Big\vert^2\cr
  &=\sum_{\alpha >0}\int\left({dk\over 2\pi}\right)^D\
  \sum_{\nu ,\nu '}\ \tilde\psi^\alpha_\nu (\uk )^*\
  H^\alpha_{\nu\nu '} (\uk )\ \tilde\psi^\alpha_{\nu '} (\uk )\ .
&(3.15)\cr}$$
Stability requires that the matrix, $H^\alpha_{\nu\nu '}(\uk )$, should be
positive definite for each root.

To obtain the frequency spectrum it is necessary to extract the
bilinear part of the Lagrangian (2.18),
$$L={\barh\over i}\ \sum_n\ \dot\phi^\mu_n\ A^j_\mu (\phi_n)
  \Lambda_{nj} -H(\phi )\ .$$
Substituting (3.8) and discarding a total derivative gives
$$\eqalignno{
  L_2 &={\barh\over 2i}\ \sum_n\ \Delta\dot\phi^\mu_n\
  \Delta\phi^\nu_n\ F^j_{\nu\mu} (\phi )\Lambda_{nj} -H_2\cr
  &=\sum_{\alpha >0}\Biggl[ {\barh\over 2i}\ \sum_n\
  \alpha\cdot\Lambda_n\ \psi^{\alpha *}_n
  ({\mathop{\partial}\limits^{\leftarrow}}_t-
  {\mathop{\partial}\limits^{\rightarrow}}_t)
  \psi^\alpha_n\ +\cr
  &\quad +{1\over 2}\ \sum_{mn}\ J_{mn}\ \Lambda_m\cdot\alpha\
  \alpha\cdot\Lambda_n\vert\psi^\alpha_n-\psi^\alpha_m\vert^2
  \Biggr]\ .
    &(3.16)\cr}$$
To obtain this expression we have used the Maurer--Cartan expression
for the curvature tensor,
$$F_{\mu\nu}\ ^j =e_\mu\ ^\alpha\ e_\nu\ ^\beta\
  c_{\alpha\beta}\ ^j\ .$$
In the Cartan--Weyl basis this gives
$$F_{\mu\nu}\ ^j\ \Lambda_{nj} =\sum_{roots}\ e_\mu\ ^\alpha\
  e_\nu\ ^{-\alpha} (\alpha\cdot\Lambda_n)\ .                     \eqno(3.17)$$
The linearized equations of motion are easily obtained,
$$-\alpha\cdot\Lambda_\nu\ {\barh\over i}\ \partial_t\ \psi^\alpha_\nu
  (\uk ) =\sum_{\nu '}\ H^\alpha_{\nu\nu '} (\uk )\
  \psi^\alpha_{\nu '} (\uk )\ .$$
The frequencies are therefore given by the zeroes of the determinant
$$\det \left(\alpha\cdot\Lambda_\nu\ \delta_{\nu\nu '}\ \barh\omega -
  H^\alpha_{\nu\nu '} (\uk )\right)\ .
\eqno(3.18)$$

To illustrate, we consider the one--dimensional chain with nearest
neighbour coupling where
there are two familiar cases, ferromagnetic and antiferromagnetic, for
which the spectrum is easily obtained.

\noindent (a)\ \ {\it Ferromagnetic ground state} $(J<0)$

In this case we have a simple lattice with $\Lambda$ independent of
$n$. The fluctuation Hamiltonian (3.13) takes the form
$$\eqalign{
  H_2 &= -J\ \sum_{\alpha >0}\ (\alpha\cdot]\Lambda )^2\ \sum_n\
  \vert\psi^\alpha_n-\psi^\alpha_{n+1}\vert^2\cr
  &=-J\ \sum_{\alpha >0}\ (\alpha\cdot\Lambda )^2\int {dk\over 2\pi}\
  \tilde\psi^\alpha (k)^*\ \vert e^{ik/2}-e^{-k/2}\vert^2\
  \psi^\alpha (k)\cr}$$
i.e.
$$H^\alpha (k) =-4J(\alpha\cdot\Lambda )^2\sin^2 k/2           \eqno(3.19)$$
so that
$$\barh\omega =-4J\ \alpha\cdot\Lambda\ \sin^2 k/2\ .    \eqno(3.20)$$
The sign of $\omega$ is not significant. If $\alpha\cdot\Lambda <0$ we
can interpret $\psi^\alpha$ as the creation operator and $\psi^{-\alpha}$
as the annihilation operator for an excitation  of energy  $\barh\vert
\omega\vert$.

\noindent (b)\ \ {\it Antiferromagnetic ground state} $(J>0)$

In this case we have two sublattices with $\Lambda_1=\Lambda =-
\Lambda_2$. The Hamiltonian (3.13) now takes the form
$$\eqalign{
  H_2 &= J\ \sum_{\alpha >0}\ (\alpha\cdot\Lambda )^2\ \sum_n\left[
  \vert\psi^\alpha_{n1}-\psi^\alpha_{n2}\vert^2 +
  \vert\psi^\alpha_{n1}-\psi^\alpha_{n-1,2}\vert^2\right]\cr
  &= J\ \sum_{\alpha >0}\ (\alpha\cdot\Lambda )^2\int {dk\over 2\pi}
  \left[\vert\psi^\alpha_1(k)-\psi^\alpha_2(k)\vert^2 +
  \vert e^{ik/2}\ \psi^\alpha_1(k) - e^{-ik/2}\
  \psi^\alpha_2(k)\vert^2\right]\cr}$$
i.e.
$$H^\alpha (k) =2J(\alpha\cdot\Lambda )^2\left[\matrix{
  1&-e^{-ik/2}\cos k/2\cr -e^{ik/2}\cos k/2 &1\cr}\right]\ .   \eqno(3.21)$$
The secular determinant (3.18) is
$$\eqalign{
  \det &\Bigg\vert\matrix{
  \alpha\cdot\Lambda\ \barh\omega -2J(\alpha\cdot\Lambda )^2
  &\ \ \ \ \ 2J(\alpha\cdot\Lambda )^2\ e^{-ik/2}\cos k/2\cr
  &\cr
  2J(\alpha\cdot\Lambda )^2\ e^{ik/2}\cos k/2
  &-\alpha\cdot\Lambda\ \barh\omega -2J(\alpha\Lambda )^2\cr}
  \Bigg\vert\cr
  &\cr
  &\quad = (\alpha\cdot\Lambda )^2\left( -\barh^2\omega^2 +4J^2
  (\alpha\cdot\Lambda )^2\sin^2 k/2\right)\cr}$$
so that
$$\barh\omega =\ \pm\ 2J(\alpha\cdot\Lambda )\sin\ k/2\ .    \eqno(3.22)$$
In contrast to the ferromagnetic case (3.20), the spectrum is linear
in $k$ near $k=0$. For symmetric spaces like $SU(N+M)/SU(N)\times
SU(M)\times U(1)$, it can be shown that $\alpha\cdot\Lambda$ is
independent of $\alpha$ and hence, so are the frequencies. The spectrum
is relativistic for small $k$. In this case the corresponding low energy
field theory will turn out to be the well known relativistic
non--linear $\sigma$--model targeted on the appropriate symmetric
space.

\SECTION{THE CONTINUUM LIMIT}

To extract a continuum description of the long wavelength behaviour
of the system one supposes that the dynamical variables are slowly
varying across the lattice,
$$\phi_m-\phi_n\ll\phi_n\ .$$
In the correspondence theory limit this means
$$\eqalignno{
  Q_\alpha (\phi_m)\ Q^\alpha (\phi_n) &\simeq \Lambda^j_m\ D_j\ ^k
  (L^{-1}_{\phi_m}\ L_{\phi_n})\Lambda_{nk}\cr
  &=\Lambda_m\cdot\Lambda_n +\cr
  &\quad +{1\over 2}\ (\phi_m-\phi_n)^\mu (\phi_m-\phi_n)^\nu\
  k^{mn}_{\mu\nu} (\phi_n) +\dots                   &(4.1)\cr}$$
where $k_{\mu\nu}$ is the tensor introduced above,
$$k^{mn}_{\mu\nu} (\phi ) =\sum_{roots}\ e_\mu\ ^\alpha
  (\phi )\ e_\nu\ ^{-\alpha} (\phi )\
  (\Lambda_m\cdot\alpha )(\alpha\cdot\Lambda_n)\ .$$

The idea is to express quantities like $H(\phi )$ as functionals of
smooth interpolating fields, $\phi (x)$,
$$H(\phi )=\int d^Dx\ {\cal H}(\phi (x),\partial\phi (x),\dots ) \eqno(4.2)$$
defining thereby a Hamiltonian density. Since the interpolating fields
are supposed to be slowly varying one aims to find the leading terms in
an expansion in powers of $\partial\phi ,\partial^2\phi$, etc. The
lattice sums one meets are usually simple enough that this expansion
can be found by inspection. However, it is probably worth pointing out
that there is a systematic procedure that employs Fourier expansions.
The lattice variables, $\phi_n$ may be represented by Fourier integrals,
$$\phi_n=\int \left({dk\over 2\pi}\right)^D\ \tilde\phi (\uk )\
  e^{i\uk\cdot\un}
    \eqno(4.3)$$
where the wave vectors are integrated over a cell of volume $(2\pi)^D$.
The Fourier components $\tilde\phi$ are periodic in $\uk$ with period
$2\pi$. The Hamiltonian may be regarded as a functional of $\tilde\phi$,
e.g.
$$H(\phi )=\sum_N\ {1\over N!}\int \left({dk_1\over 2\pi}\right)^D
  \dots \left({dk_N\over 2\pi}\right)^D\ H_N (k_1\dots k_N)\
  \tilde\phi (k_1)\dots\tilde\phi (k_N)
\eqno(4.4)$$
where the coefficient functions, $H_N$, are of course periodic in the
wave vectors $\uk_1,\uk_2,\dots\uk_N$. If $\phi_n$ is slowly varying
then $\tilde\phi (\uk )$ is non--vanishing only in the neighbourhood of
$\uk =0$ (mod $2\pi$). In this case it is therefore legitimate to expand
the coefficient functions in powers of $\uk_1,\dots ,\uk_N$, e.g.
$$H_N(k_1\dots k_N)=(2\pi )^D\ \delta (\Sigma k)\left( h^{(0)}_N-
  \Sigma\ \uk_j\cdot\uk_\ell\ h^{(2)}_N+\dots\right)\ .$$
The periodicity in $\uk$ is now irrelevant and one may substitute
$$(2\pi )^D\ \delta (\Sigma k)=\int d^Dx\ e^{i\Sigma\uk\cdot\ux}\ .$$
Define the interpolating fields $\phi (x)$ by the continuum Fourier
integral,
$$\phi (x) =\int\left({dk\over 2\pi}\right)^D\ \tilde\phi (\uk )\
  e^{i\uk\cdot\ux}\ .
\eqno(4.5)$$
One then obtains $H(\phi )$ in the form (4.2) with
$$\eqalignno{
  {\cal H} &=\sum_N\ {1\over N!}\left[ h^{(0)}_N\ \phi (x)^N +
  \left({N\atop 2}\right)\ h^{(2)}_N\ \phi (x)^{N-2}\ (\upar\phi )^2
  +\dots\right]\cr
  &= V(\phi )+Z(\phi ) (\upar\phi )^2 +\dots\ .              &(4.6)\cr}$$

In the following we shall discard terms containing more than two
derivatives of the interpolating fields. Since the expression (4.1) is
bilinear in $\phi_m-\phi_n$ it will be sufficient to keep only the first
derivative in this quantity. However, a minor complication arises from
the need to keep track of the various translation invariant sublattices.
We write
$$\phi_{\un\nu} =\phi_{\un} +\xi_{\un\nu}                          \eqno(4.7)$$
where $\un$ is a $D$--vector with integer--valued components and
$\nu =1,\dots ,f$ labels the sublattice. The variables $\xi_{\un\nu}$
are constrained to satisfy
$$\sum_\nu\ \xi_{\un\nu} =0
\eqno(4.8)$$
and they are small quantities of order, $\phi_{\un}-\phi_{\um}$.
Hence, to leading order,
$$\phi_{\un\nu} -\phi_{\un '\nu '} =\left[
  (\un -\un ')\cdot\upar\phi (x) +\xi_\nu (x) -\xi_{\nu '}(x)
  \right]_{\ux =\un}
    \eqno(4.9)$$
which is to be substituted into the formula for the Hamiltonian,
$$\eqalign{
  H(\phi ) &={1\over 2}\ \sum_{n\nu n'\nu '}\ J_{\un\nu ,\un '\nu '}
  \Biggl[ \Lambda_\nu\cdot\Lambda_{\nu '} +\cr
  &\quad +{1\over 2} (\phi_{\un\nu} -\phi_{\un '\nu '})^\sigma
  (\phi_{\un\nu}-\phi_{\un '\nu '})^\rho\
  k^{\nu\nu '}_{\sigma\rho} (\phi_{\un}) +\dots\Biggr]\cr}$$
Using the translation invariance of $J_{\un\nu ,\un '\nu '}$, one obtains
the density,
$$\eqalignno{
  {\cal H} &= {1\over 2}\ \sum_{\un\nu\nu '}\ J_{\un\nu ,\uo\nu '}
  \Biggl[ \Lambda_\nu\cdot\Lambda_{\nu '} +\cr
  &\quad + {1\over 2} (\un\cdot\upar\phi +\xi_\nu -\xi_{\nu '}
  )^\sigma (\un\cdot\upar\phi +\xi_\nu -\xi_{\nu '})^\rho\
  k^{\nu\nu '}_{\sigma\rho} (\phi )+\dots\Biggr]          &(4.10)\cr}$$
It is generally assumed that the coupling strengths fall off rapidly with
distance between sites so that the sum over $\un$ in this expression
contains only a few terms. Defining the moments,
$$\eqalignno{
  \sum_{\un}\ J_{\un\nu ,\uo\nu '} &= J_{\nu\nu '}=J_{\nu '\nu }\cr
  \sum_{\un}\ n^i\ J_{\un\nu ,\uo\nu '} &= J^i_{\nu\nu '} =-
  J^i_{\nu '\nu}\cr
  \sum_{\un}\ n^in^j\ J_{\un\nu .\uo\nu '} &= J^{ij}_{\nu\nu '} =
  J^{ij}_{\nu '\nu}
&(4.11)\cr}$$
one obtains
$$\eqalignno{
  {\cal H} &= {1\over 2}\ \sum_{\nu\nu '} \Biggl[ J_{\nu\nu '}\
  \Lambda_\nu\cdot\Lambda_{\nu '} +\cr
  &+{1\over 2} \Bigl( J^{ij}_{\nu\nu '}\ \partial_i\phi^\rho\
  \partial_j\phi^\sigma +2J^i_{\nu\nu '}\ \partial_i\phi^\rho
  (\xi_\nu -\xi_{\nu '})^\sigma +\cr
  &\quad + J_{\nu\nu '} (\xi_\nu -\xi_{\nu '} )^\rho
  (\xi_\nu -\xi_{\nu '})^\sigma\Bigr)
  k^{\nu\nu '}_{\rho\sigma}(\phi )+\dots\Biggr]           &(4.12)\cr}$$
Although this general result may look rather forbidding, it can simplify
in specific applications. For example, if the couplings are isotropic and
$J^i=0, J^{ij}\sim\delta^{ij}$. In the ferromagnetic case there are no
sublattices so the variables $\xi_\nu$ disappear along with the
label, $\nu$. In the antiferromagnetic case there are usually only two
sublattices and $\xi_2=-\xi_1,\Lambda_2=-\Lambda_1$ and, moreover,
one usually assumes $J_{11}=J_{22}=0$, etc.

To complete the discussion of the continuum limit, it is necessary to
consider the kinetic term,
$${\barh\over i}\ \sum_{\un\nu}\ \dot\phi^\mu_{\un\nu}\
  A^j_\mu (\phi_{\un\nu} )\Lambda_{\nu j}                         \eqno(4.13)$$
where $A_\mu$ is the spin connection on $G/H$. Substituting the
expression (4.7) for $\phi_{\un ,\nu}$ and expanding in powers of $\xi$,
one obtains after discarding a total derivative,
$$\eqalignno{
  & (\dot\phi +\dot\xi_\nu )^\rho\ A_\rho (\phi +\xi_\nu ) =\cr
  &\quad =\dot\phi^\rho\ A_\rho (\phi )+\xi^\rho_\nu
  \left(\dot\phi^\lambda +{1\over 2}\ \dot\xi^\lambda_\nu
  \right)\ F_{\rho\lambda} +{1\over 2}\ \xi^\rho_\nu\ \xi^\sigma_\nu\
  \dot\phi^\lambda\ \partial_\sigma\ F_{\rho\lambda} +\dots &(4.14)\cr}$$
where $A_\rho =A^j_\rho\ \Lambda_{\nu j}$ and $F_{\rho\lambda}$ is the
corresponding curvature tensor.

In the ferromagnetic case there is nothing further to do. Only the first
term in (4.14) is present and the equations of motion will imply
$\dot\phi /\phi\sim O(k^2)$ to the approximation used in (4.12). In
other cases, if we impose the condition,
$$\sum_\nu\ \Lambda_{\nu j} =0\ ,
\eqno(4.15)$$
then the first erm in (4.14) is absent and the most important term
becomes
$${\barh\over i}\ \sum_\nu\ \xi^\rho_\nu\ \dot\phi^\lambda\
  F^j_{\rho\lambda}\ \Lambda_{\nu j}\ .
\eqno(4.16)$$
Here the equations of motion imply that both $\dot\phi /\phi$ and
$\xi_\nu$ are $O(k)$. It is therefore consistent in this approximation
to discard all other terms in (4.14). Notice, in particular, that
$\dot\xi_\nu$ disappears from the Lagrangian. This means that $\xi_\nu$
becomes an algebraic variable and can be eliminated from the
dynamics \FOOTNOTE{$^{*)}$}{If the condition (4.15) is not imposed then
$\xi_\nu$ becomes a true dynamical variable associated with optical
branches in the spectrum, i.e. $\dot\xi /\xi$ remains finite in the
limit $k\to 0$.}.

To conclude, the Lagrangian for the ferromagnetic spin waves is given by
$$L = \int d^Dx\Biggl[ {\barh\over i}\ \partial_t\ \phi^\mu\
  A_\mu (\phi )-
  {1\over 4}\ J^{ij}\ \partial_i\ \phi^\mu\
  \partial_j\phi^\nu\ k_{\mu\nu} (\phi )\Biggr]            \eqno(4.17)$$
where $A_\mu =A^j_\mu\ \Lambda_j$ and
$$k_{\mu\nu} (\phi ) =\sum_{roots}\ e_\mu\ ^\alpha\
  e_\nu\ ^{-\alpha}\ (\alpha\Lambda )^2\ .$$
The tensor $J^{ij}$ is assumed to be negative definite.

The equations of motion derived from (4.17) are generalizations of the
well known Landau--Lifshitz $^{13)}$ equations for ferromagnetic
spin waves. They are
$${\barh\over i}\ F_{\mu\nu}\ \partial_t\ \phi^\nu = {J\over 2}\
  k_{\mu\nu}\ \Delta\ \phi^\nu$$
where we have assumed an isotropic coupling, i.e.
$$J^{ij}=J\ \delta^{ij}\quad J<0$$
and where the ``Laplacian'' is given by
$$\Delta\ \phi^\nu=\partial^2_i\ \phi^\nu +{1\over 2}\ k^{\nu\rho}
  \left(\partial_\lambda\ k_{\rho\mu}+\partial_\mu\ k_{\rho\lambda}-
  \partial_\rho\ k_{\lambda\mu}\right)\ \partial_i\ \phi^\lambda\
  \partial_i\ \phi^\mu\ .$$
It is not difficult to verify that for $G/H=SU(2)/U(1)$ these equations
assume their familiar form~$^{13)}$
$$\barh\ \partial_t\ \vec n=-{J\over 2}\ s\ \vec n\times
  \nabla_i\ ^2\ \vec n$$
where $s$ is the spin and $\vec n$ is a unit vector describing the
target space $S^2$.

For the antiferromagentic case (with two sublattices),
$$\eqalign{
  \Lambda_1 &= \Lambda =-\Lambda_2\cr
  \xi_1 &= \xi =-\xi_2\cr}$$
the Lagrangian is given by
$$\eqalignno{
  L &= \int d^Dx\Biggl[ {\barh\over i}\ 2\xi^\mu\ F_{\mu\nu}\
  \partial_t\phi^\nu -\cr
  &\qquad -\left( J^{ij}\ \partial_i\phi^\mu\ \partial_j\phi^\nu +
  4J^i\ \partial_i\phi^\mu\ \xi^\nu +4J\ \xi^\mu\xi^\nu\right)
  k_{\mu\nu} (\phi )\Biggr]
&(4.18)\cr}$$
where, for simplicity, we have taken $J^{ij}_{11}=J^{ij}_{22}=0$ and
we have suppressed the sublattice indices, writing
$$J^{ij}_{12} =J^{ij}_{21} =J^{ij},\qquad
   J^i_{12} =-J^i_{21} =J^i$$
$$J_{12} =J_{21}=J$$
and
$$k^{12}_{\mu\nu} =k^{21}_{\mu\nu} =k_{\mu\nu} =-
  \sum_{roots}\ e_\mu\ ^\alpha\ e_\nu\ ^{-\alpha}
  (\alpha\Lambda )^2\ .
\eqno(4.19)$$
The tensor $J^{ij}$ is assumed to be positive definite. The auxiliary
variable $\xi^\mu$ can be eliminated from (4.18) by solving the Euler
Lagrange equation $\delta L/\delta\xi^\mu =0$,
$${\barh\over i}\ F_{\mu\nu}\ \dot\phi^\nu -2k_{\mu\nu}
  (J^i\ \partial_i\phi^\nu +2J\ \xi^\nu )=0\ .$$
This gives
$$\eqalignno{
  L &=\int d^Dx\Biggl[ {\barh^2\over 4J}\ g_{\mu\nu}\ \partial_t
    \phi^\mu\ \partial_t\phi^\nu -{\barh\over i}
    {J^i\over J}\ \partial_i\phi^\mu\ F_{\mu\nu}\ \partial_t
    \phi^\nu\cr
    &\qquad -\left( J^{ij}-{J^iJ^j\over J}\right)
    \partial_i\phi^\mu\ \partial_j\phi^\nu\ k_{\mu\nu}
    \Biggr]
       &(4.20)\cr}$$
where the tensor $g_{\mu\nu}(\phi )$ is defined by
$$\eqalignno{
  g_{\mu\nu} &= -F_{\mu\lambda}\ F_{\nu\rho}\ (k^{-1})^{2\rho}\cr
  &=-\sum_{roots}\ e_\mu\ ^\alpha\ e_\nu\ ^{-\alpha}      &(4.21)\cr}$$
(The formulae (4.19) and (3.17) have been used.) The tensor $g_{\mu\nu}$
is $G$--invariant and positive definite.

For symmetric spaces there is a unique $G$--invariant second rank
symmetric tensor on $G/H$ implying that
$$k_{\mu\nu} (\phi ) =v^2\ g_{\mu\nu} (\phi )$$
when $v$ is independent of $\phi$. If in addition we also have
invariance under space rotations, then
$$J^{ij} -{J^iJ^j\over J} \sim \delta^{ij}$$
and the theory becomes fully relativistic. This obtains automatically
for $D=1$ for which we recover the well known two--dimensional
$\sigma$--models targeted on symmetric spaces.

The Lagrangian (4.20) gives a sensible description of antiferromagnetic
spin waves if
$$J>0,\quad J^{ij} -{J^iJ^j\over J} >0\ .
\eqno(4.22)$$
The middle term in (4.20),
$$-{\barh\over i}\ {J^i\over J}\ \partial_i\phi^\mu\
  F_{\mu\nu}\ \partial_t\phi^\nu
\eqno(4.23)$$
is of a topological nature. It makes no contribution to the classical
equations of motion.
Indeed, it can be expressed as a total derivative. The integral of this
quantity over a spacetime region, $M$, reduces to a boundary integral,
$$\eqalignno{
  &-{\barh\over i}\ {J^i\over J}\int_M dt\ d^Dx\ \partial_i\phi^\mu\
  \partial_t\phi^\nu\ F_{\mu\nu} =\cr
  &\qquad =-{\barh\over i}\ {J^i\over J}\int_M dt\ d^Dx
  \left[ \partial_i (\partial_t\phi^\mu\ A_\mu ) -\partial_t
  (\partial_i\phi^\mu\ A_\mu )\right]\ .                        &(4.24)\cr}$$
If $M$ is a compact two--dimensional manifold this integral is a
topological invariant -- an integer multiple of $2\pi\barh$ for the
2--sphere as will be shown in the following section. In general the
integral (4.24) has no topological significance. For compact manifolds
of higher dimension it has been argued that this term should vanish
$^{10)}$.

It was shown by Haldane, who discovered the term in the case of a
one--dimensional antiferromagnetic spin system with nearest
neighbour couplings, that the topological contribution gives an
alternating sign in the sum over configurations for half--integer spin
systems $^{9)}$. In the following we consider some classical aspects
of the one--dimensional case.

\SECTION{INSTANTONS}

The continuum theory that emerges from the one--dimensional
antiferromagnetic chain is described by a Lagrangian of the type
(4.20). Consider now the Euclidean version of this theory. The
Lagrangian density is
$${\cal L} ={1\over 4J}\ g_{\mu\nu}\ \partial_t\phi^\mu\
  \partial_t\phi^\nu +J'\ k_{\mu\nu}\ \partial_x\phi^\mu\
  \partial_x\phi^\nu
\eqno(5.1)$$
where we take $\barh =1$ and discard the topological term. The
tensors $g_{\mu\nu}$ and $k_{\mu\nu}$, given by (4.21) and (4.19),
respectively, are positive definite and we shall assume that $J$ and $J'$
are positive.

There is a simple argument that suggests the possible existence of
finite action solutions of the equations of motion, i.e. the instantons.
Consider the expressions
$$\eqalignno{
  &{1\over 4J}\ g_{\mu\nu}\left(\partial_t\phi^\mu\pm\ 2i\sqrt{JJ'}\
  F^\mu\ _\lambda\ \partial_x\phi^\lambda\right)
  \left(\partial_t\phi^\nu\pm\ 2i\sqrt{JJ'}\ F^\nu\ _\rho\
  \partial_x\phi^\rho\right)\cr
  &\quad ={1\over 4J}\ g_{\mu\nu}\ \partial_t\phi^\mu\
  \partial_t\phi^\nu\pm i\sqrt{{J'\over J}}\ F_{\nu\lambda}\
  \partial_x\phi^\lambda\ \partial_t\phi^\nu\cr
  &\qquad -J'\ g_{\mu\nu}\ F^\mu\ _\lambda\ F^\nu\ _\rho\
  \partial_x\phi^\lambda\ \partial_x\phi^\rho\cr
  &\quad ={1\over 4J}\ g_{\mu\nu}\ \partial_t\phi^\mu\
  \partial_t\phi^\nu\pm\ i\sqrt{{J'\over J}}\ F_{\nu\lambda}\
  \partial_x\phi^\lambda\ \partial_t\phi^\rho + J'k_{\mu\nu}\
  \partial_x\phi^\mu\ \partial_x\phi^\nu                    &(5.2)\cr}$$
where
$$\eqalignno{
  g_{\mu\nu}\ F^\nu\ _\rho &= F_{\mu\rho}\cr
  &=\sum_{roots}\ e_\mu\ ^\alpha\ e_\rho\ ^{-\alpha}\
  \alpha\cdot\Lambda\ .                                           &(5.3)\cr}$$
In a basis of real coordinates the components $F_{\mu\nu}$ are pure
imaginary. Hence the expresssions (5.2) are real and positive, i.e.
$$\eqalign{
  {\cal L} &={1\over 4J}\ g_{\mu\nu}\ \partial_t\phi^\mu\
  \partial_t\phi^\nu +J'\ k_{\mu\nu}\ \partial_x\phi^\mu\
  \partial_x\phi^\nu\cr
  &\geq\Bigg\vert\sqrt{{J'\over J}}\ F_{\mu\nu}\ \partial_x\phi^\mu\
  \partial_t\phi^\nu\Bigg\vert\ .\cr}$$
This means that the Euclidean action has a lower bound,
$$\eqalignno{
  \int d^2x &\left[ {1\over 4J}\ g_{\mu\nu}\ \partial_t\phi^\mu\
  \partial_t\phi^\nu +J'\ k_{\mu\nu}\ \partial_x\phi^\mu\
  \partial_x\phi^\nu\right]\cr
  &\quad\geq\Bigg\vert\sqrt{{J'\over J}}\ \int d^2x\ F_{\mu\nu}\
  \partial_x\phi^\mu\ \partial_t\phi^\nu\Bigg\vert\ .    &(5.4)\cr}$$
It remains to determine whether there are configurations for which
the integral on the right--hand side is finite.

Suppose that the two--dimensional Euclidean spacetime has the
topology of a sphere. The coordinate functions, $\phi^\mu (x)$, define the
image of this sphere in the manifold, $G/H$. It is not difficult to see that
such maps have a topological classification, The manifold is generally
covered by more than one coordinate patch. If the coordinates in
two patches that intersect are associated with the group elements
$L^{(1)}_\phi$ and $L^{(2)}_\phi$, as described above in Sec.2, then at
points in the overlap there is a relation,
$$L^{(2)}_\phi =L^{(1)}_\phi\ h^{(1,2)} (\phi )
\eqno(5.5)$$
where $h^{(1,2)}\in H$. Only the Abelian part of $h^{(1,2)}$ is of interest
because this is what determines the relation between $A^j(\phi )$ in the
two patches. With
$$h^{(1,2)} (\phi ) =e^{i\psi^j(\phi )H_j}
\eqno(5.6)$$
it follows from the definition (2.12) that
$$A^j_{(2)}=A^j_{(1)}-i\ d\psi^j\ .
\eqno(5.7)$$
If the overlap between the two patches is not simply connected then the
angles $\psi^j(\phi )$ can be multiple valued, i.e.
$$\Delta\psi^j =\oint\ d\psi^j\not= 0$$
where the integral is taken around a closed path in the overlap.
However, it is essential that the group element $h^{(1,2)}$ be single
valued and this implies
$$\Delta\psi^j\ \Lambda_j\in 2\pi\bbz\ .
\eqno(5.8)$$
On the other hand, the flux of $F^j\ \Lambda_j$ associated with a
compact 2--space must reduce to a sum of such terms. With the
2--sphere suppose that the northern hemisphere maps into patch 1 while the
southern maps into patch 2. Then the flux is given by
$$\eqalignno{
  \int F^j\ \Lambda_j &= \int dA^j\ \Lambda_j\cr
  &=\oint (A^j_{(1)}-A^j_{(2)})\Lambda_j\cr
  &=i\ \Delta\psi^j\ \Lambda_j
&(5.9)\cr}$$
where the contour runs around the image of the equator. The flux is an
integer multiple of $2\pi i$. This integer, $N$, serves to classify the
configurations and, in each class the Euclidean action is bounded below
by $\vert 2\pi N\sqrt{J'/J}\ \vert$.

To saturate the bound one needs to solve the first order differential
equations,
$$g_{\mu\nu}\ \partial_t\phi^\nu +2i\sqrt{JJ'}\ F_{\mu\nu}\
  \partial_x\phi^\nu =0$$
or, more explicitly,
$$(\partial_t\phi^\mu -2i\sqrt{JJ'}\ \alpha\cdot\Lambda\
  \partial_x\phi^\mu )e_\mu\ ^\alpha =0\ .
\eqno(5.10)$$
The solutions of these equations comprise the instantons. If $G/H$ is
a complex symmetric space it can be shown that Eq.(5.10) reduces to
the well studied case $^{12)}$
$$\partial_{\bar z}\ \zeta^i=0$$
where $z$ is a complex coordinate in the 2--dimensional Euclidean
spacetime and $(\zeta^i,\bar\zeta^i)$ is a set of complex coordinates
on the target space $G/H$.

Returning to Haldane's observation, it will be seen that our result (4.20)
appears to differ in some respects. The Euclidean version of (4.20)
with topological term included is
$$\eqalignno{
  {\cal L} &={1\over 4J}\ g_{\mu\nu}\ \partial_t\phi^\mu\
  \partial_t\phi^\nu +{J^x\over J}\ F_{\mu\nu}\ \partial_x\phi^\mu\
  \partial_t\phi^\nu +\cr
  &\quad +\left( J^{xx}-{J^xJ^x\over J}\right) k_{\mu\nu}\
  \partial_x\phi^\mu\ \partial_x\phi^\nu
&(5.11)\cr}$$
where we have again set $\barh =1$. The coefficients $J,J^x$ and
$J^{xx}$ are given in terms of the spin couplings, $J_{n\sigma ;n'
\sigma '}$, by the lattice sums (4.11). In particular,
$$J=\sum_n\  J_{n,1; o,2}\quad {\rm and}\ J^x=\sum_n\ n\
  J_{n,1;o,2}\ .
      \eqno(5.12)$$
Due to the presence of the topological term the Euclidean action
contains the imaginary contribution
$$2\pi i\ N\ J^x/J$$
where $N$ is the instanton number. Haldane's result, obtained for
$SU(2)$ spin models with nearest neighbour couplings, is simply
$i\pi N$. In fact, this result does agree with ours since it can be
shown quite generally that $J^x=J/2$. The relevant coupling parameters
are
$$\eqalignno{
  J_{n,1;n,2} &=J_{n+1,1;n2} = K_1,\cr
  J_{n-1,1;n2} &=J_{n+2,1;n2} =K_3,\cr
  J_{n-2,1;n2} &= J_{n+3,1;n2} =K_5,                   &(5.13)\cr}$$
etc., corresponding to nearest, third nearest, fifth nearest, . . .
neighbours on the lattice. Substituting into (5.12) gives
$$\eqalignno{
  J &= 2K_1+2K_3+2K_5 +\dots\cr
  J^x &=K_1+K_3+K_5 +\dots                                &(5.14)\cr}$$
which confirms Haldane's result in the more general context.

To summarize, the Euclidean path integral for the generalized
antiferromagnetic spin chains includes the sign factor, $e^{i\pi N}$,
where $N$ is an integer defined by the flux integral,
$$\eqalignno{
  N &= {1\over 2\pi}\int F^j\ \Lambda_j\cr
  &={1\over 2\pi}\int F_j\ \Lambda^j\cr
  &=N_j\ \Lambda^j\ .                                      &(5.15)\cr}$$
The weights $\Lambda^j=g^{jk}\ \Lambda_k$ are integers for any
finite dimensional unitary representation of $G$. The coefficients,
$N_j$ are integers that characterize the configuration, $\phi^\mu (x)$.
Models in which the $\Lambda^j$ are all even will not be sensitive to
the classes of configurations -- like the integer spin $SU(2)$ models.
Other models, in which one or more of the $\Lambda^j$ are odd, will be
sensitive -- they must generalize, in various ways, the half--integer
spin $SU(2)$ models.

\SECTION{OPERATOR REALIZATION FOR $SU(N+1)$}

To resolve the factor ordering ambiguities in the quantized theory it is
necessary to construct a realization of the generators $Q_\alpha (\phi )$
in terms of the dynamical variables $\phi_n$. The details of such a
realization will depend on the nature of the target manifold $G/H$
and the quantum numbers of the reference state, $\vert\Lambda >$.
We are not able to give a general solution but we can illustrate a
method which works in at least some cases. Here we discuss the
realization of $SU(N+1)$ in terms of operators associated with the
coordinates of $\bbc P^N$.

The algebra of $SU(N+1)$ is spanned by the $N(N+2)$ generators,
$Q_A\ ^B, A,B=1,2,\dots ,N+1$, which satisfy the commutation rules
$$[Q_A\ ^B, Q_C\ ^D]=\delta^D_A\ Q_C\ ^B-\delta^B_C\
  Q_A\ ^D
         \eqno(6.1)$$
and the constraint, $Q_A\ ^A =0$. The fundamental representation
of this algebra is generated by the $(N+1)$--dimensional traceless
hermitian matrices,
$$(Q_A\ ^B)_C\ ^D =\delta^D_A\ \delta^B_C -
  {1\over N+1}\ \delta^B_A\ \delta^D_C\ .
\eqno(6.2)$$

We shall consider cosets of $SU(N+1)$ with respect to the subgroup
$SU(N)\times U(1)$ generated by the $N^2$ operators, $Q_i\ ^j, i,j=1,\dots ,
N$. The finite transformations, $L_\phi$, that represent the cosets can be
chosen such that, in the fundamental representation,
$$(L_\phi )_C\ ^D =\left[\matrix{
  \sqrt{1-\phi\phi^+} &\phi\cr &\cr
  -\phi^+ & \sqrt{1-\phi^+\phi}\cr}\right]
\eqno(6.3)$$
where $\phi$ is an $N$--component column vector. The coordinates,
$\phi_i$, on the manifold, $\bbc P^N$, belong to the fundamental
representation of $SU(N)$. It is not difficult to verify that the matrix
(6.3) is both unitary and unimodular.

According to the general discussion of Sec.2, the reference state,
$\vert\Lambda >$, from which the coherent states are generated, must
be invariant up to a phase under the action of the stability group,
$SU(N)\times U(1)$. In particular, it must be a singlet of $SU(N)$.
We are therefore restricted to those representations of $SU(N+1)$ which
contain such a singlet. In terms of Young tableaux these representations
$(n_1,n_2)$ are characterized by $n_1$ columns with $N$ boxes and
$n_2$ columns with 1 box. Each of these representations contains just
one singlet of $SU(N)$. We shall find that the realizations to be
obtained depend on $n_1-n_2$, only.

The Cartan subalgebra of $SU(N+1)$ is spanned by the $N$ elements,
$Q_1\ ^1, Q_2\ ^2,\dots , Q_N\ ^N$, whose eigenvalues define the
components of the weight vectors
$$Q_j\ ^j\ \vert\Lambda >\ = \vert\Lambda >\ \Lambda_j,\qquad
  j=1,\dots , N\ .$$
Since the reference state is a singlet of $SU(N)$ these components are
all equal
$$\Lambda_1=\Lambda_2=\dots =\Lambda\ .                         \eqno(6.4)$$
It is convenient to define the hypercharge operator
$$Y=Q_{N+1}\ ^{N+1} =-\sum^N_1\ Q_j\ ^j
\eqno(6.5)$$
to stand for the $SU(N)$ singlet generator. In the fundamental
representation it is a diagonal matrix,
$$Y_C\ ^D =\left[\matrix{ -1/(N+1) &&&\cr &\ddots&&\cr
  &&-1/(N+1) &\cr &&&N/(N+1)\cr}\right]\ .
\eqno(6.6)$$
The value of the hypercharge in the reference state is given by
$$Y\vert\Lambda >\ = -\sum_j\ Q_j\ ^j\ \vert\Lambda >\ =
  -N\Lambda\vert\Lambda >\ .$$
On the other hand, since this state is the singlet component of a
representation contained in the direct product of $n_2$ fundamental
representations and $n_1$ conjugates, it follows that
$$Y=(n_2-n_1)\ {N\over N+1}$$
or
$$\Lambda = {n_1-n_2\over N+1}\ .
\eqno(6.7)$$

The $U(1)$--component of the spin connection on $\bbc P^N$ is defined by
$$L^{-1}_\phi\ dL_\phi =i\ A\ Y +\dots$$
With the parametrization chosen in (6.3) this gives
$$A={1\over i}\ {N+1\over 2N}\ (\phi^+\ d\phi -d\phi^+\phi )\ .  \eqno(6.8)$$
The overlap between neighbouring coherent states is therefore,
$$\eqalign{
  <\phi +d\phi\vert\phi >\ &=\ <\Lambda\vert L^{-1}_{\phi +d\phi}\
  L_\phi\ \vert\Lambda >\cr
  &= 1-i\ A\ <\Lambda\vert Y\vert\Lambda >\cr
  &= 1+{1\over 2}\ (N+1)\Lambda (\phi^+\ d\phi -
  d\phi^+\phi )\cr}$$
where $\Lambda$ is given by (6.7). Hence the Lagrangian (2.18) in this
case takes the form
$$L={\barh\over 2i}\ (N+1)\Lambda (\phi^+\ \partial_t\phi -
  \partial_t\phi^+\phi )-H(\phi ,\phi^+ )\ .
\eqno(6.9)$$
Canonical quantization gives the commutation rules
$$\eqalignno{
  [\phi_i,\phi_j] &= 0\cr
  [\phi_i,\phi^+_j] &= -{1\over (N+1)\Lambda}\ \delta_{ij}\cr
  [\phi^+_i,\phi^+_j] &= 0
&(6.10)\cr}$$
where $(N+1)\Lambda$ is an integer.

To construct the generators out of these operators it is helpful to
consider firstly their coherent state expectation values,
$$\eqalign{
  <\phi\vert Q_A\ ^B\ \vert\phi >\ &= \ <\Lambda\vert\ L_\phi^{-1}\
  Q_A\ ^B\ L_\phi\ \vert\Lambda >\cr
  &=(L_\phi )_A\ ^C\ <\Lambda\vert\ Q_C\ ^D\ \vert\Lambda >\
  (L^{-1}_\phi )_D\ ^B\cr
  &=-(N+1)\Lambda\ (L_\phi\ Y\ L^{-1}_\phi )_A\ ^B\cr}$$
where $Y$ is the hypercharge matrix in the fundamental representation
(6.6). With the matrices $(L_\phi )_A\ ^B$ given by (6.3) one obtains,
$$\eqalignno{
  <\phi\vert\ Q_A\ ^B\ \vert\phi >\ &=\left[\matrix{
  \Lambda -(N+1)\Lambda\ \phi\phi^+
  &-(N+1)\Lambda\sqrt{1-\phi^+\phi}\ \phi\cr
  &\cr
  -(N+1)\Lambda\sqrt{1-\phi^+\phi}\ \phi^+
  &-N\Lambda +(N+1)\Lambda\ \phi^+\phi\cr}\right]            &(6.11)\cr}$$
using the elementary identities
$$\sqrt{1-\phi\phi^+}\ \phi =\phi\sqrt{1-\phi^+\phi}$$
and
$$\phi^+\sqrt{1-\phi\phi^+} =\sqrt{1-\phi^+\phi}\ \phi^+\ .$$
The classical expressions on the right--hand side of (6.11) suggest the
following operator realization for $\Lambda <0$,
$$\eqalignno{
  Q_i\ ^j(\phi ) &=\Lambda\ \delta^j_i-(N+1)\Lambda\
  \phi^+_j\ \phi_i\cr
  &=(\Lambda -1)\delta^j_i-(N+1)\Lambda\ \phi_i\ \phi^+_j\cr
  Q_i\ ^{N+1}(\phi ) &=-(N+1)\Lambda\ \sqrt{1-\phi^+\phi}\ \phi_i\cr
  &=-(N+1)\Lambda\ \phi_i\ \sqrt{1-{1\over (N+1)\Lambda} -
  \phi^+\phi}\cr
  Q_{N+1}\ ^j(\phi ) &=-(N+1)\Lambda\ \phi^+_j\ \sqrt{1-\phi^+\phi}\cr
  &=-(N+1)\Lambda\ \sqrt{1-{1\over (N+1)\Lambda} -\phi^+\phi}\
  \phi^+_j\cr
  Q_{N+1}\ ^{N+1}(\phi ) &=-N\Lambda +(N+1)\Lambda\ \phi^+\phi\ .

          &(6.12)\cr}$$
It is easy to verify, using the commutation rules (6.10) that these
operators satisfy the algebra of $SU(N+1)$. The reference state is
annihilated by $\phi_i$. The basis for a finite dimensional unitary
representation of $SU(N+1)$ is given by
$$\vert\Lambda >,\quad \phi^+_i\ \vert\Lambda >,\quad
  \phi^+_{i_1}\ \phi^+_{i_2}\ \vert\Lambda >,\dots ,
  \phi^+_{i_1}\dots \phi^+_{i_p}\ \vert\Lambda >             \eqno(6.13)$$
where $p=-(N+1)\Lambda =n_2-n_1$ is a positive integer. The rest of
the Fock space, spanned by vectors with $p+1,p+2,\dots $ creation
operators, carries a non--unitary infinite dimensional representation
of $SU(N+1)$.

Similarly, for $\Lambda >0$ the operator realization is given by
$$\eqalignno{
  Q_i\ ^j(\phi ) &=\Lambda\ \delta^j_i-(N+1)\Lambda\ \phi_i\
  \phi^+_j\cr
  &=(\Lambda +1)\delta^j_i-(N+1)\Lambda\ \phi^+_j\ \phi_i\cr
  Q_i\ ^{N+1}(\phi ) &=-(N+1)\Lambda\ \phi_i\
  \sqrt{1+{N\over (N+1)\Lambda} -\phi^+\phi}\cr
  &=-(N+1)\Lambda\ \sqrt{1+{1\over\Lambda} -\phi^+\phi}\ \phi_i\cr
  Q_{N+1}\ ^j(\phi ) &=-(N+1)\Lambda\ \sqrt{1+{N\over (N+1)\Lambda} -
  \phi^+\phi}\ \phi^+_i\cr
  &=-(N+1)\Lambda\ \phi^+_i\ \sqrt{1+{1\over\Lambda} -
  \phi^+\phi}\cr
  Q_{N+1}\ ^{N+1}(\phi ) &=-(\Lambda +1)N+(N+1)\Lambda\ \phi^+\phi\ .

             &(6.14)\cr}$$
Here the reference state is annihilated by $\phi_i^+$ and the basis for
a finite dimensional unitary representation is given by
$$\vert\Lambda >,\quad \phi_i\vert\Lambda >,\quad
  \phi_{i_1}\phi_{i_2}\ \vert\Lambda >,\dots ,
  \phi_{i_1}\dots\phi_{i_p}\ \vert\Lambda >
\eqno(6.15)$$
where $p=(N+1)\Lambda =n_1-n_2$.

It appears from the structure of the basis sets, (6.13) and (6.15) that
the operators (6.12) and (6.14) realize, respectively, the unitary
representations $(0,n_2-n_1)$ and $(n_1-n-2,0)$ but not the general
case $(n_1,n_2)$. Our approach clearly leaves something to be desired.
This should not be a surprise since we began with the assumption that
commutators should be ignored in the first approximation -- an
assumption that is appropriate to the correspondence theory limit.
The resulting formula should therefore be read as corrections to this
limiting case valid, we expect, for large $n_1$ and $n_2$ such that
$$\vert n_1-n_2\vert\ll n_1+n_2\ .$$
We believe that it should be possible to construct exact expressions from
$\phi$ and $\phi^+$ that realize the general case $(n_1,n_2)$ and which
reduce to the forms (6.12), (6.14) when higher powers of
$\vert n_1-n_2\vert /(n_1+n_2)$ are neglected. Probably these
expressions will have branch point singularities at $\phi =0$.
In support of the conjecture we have examined a
Holstein--Primakoff type realization for the $SU(2)$ case which depends
explicitly on the spin, $s$, and helicity, $\lambda$, and which reduces
to the standard Holstein--Primakoff form when $\lambda =\pm\ s$.
This realization is discussed in the appendix.

Finally, in applying these formulae to an $SU(N+1)$ lattice model one
must construct the Hamiltonian operator, $H(\phi ,\phi^+)$ and set up the
Feynman rules. Typically, the free propagator, $<T\ \phi_n\ \phi^+_m>$,
will have the Fourier representation
$$G(\omega ,k)={1\over (n_1-n_2)\barh\omega -H(k)}$$
and the higher order contributions will be suppressed by powers of
$(n_1-n_2)^{-1}$. Corrections to the classical theory are obtained in
the usual way as a loop expansion. Of course, at any finite order the
approximate charge operators will give rise to transitions into the
unphysical (non--unitary) part of the representation. This defect is
endemic to the semiclassical treatment of spin waves.

\SECTION{OUTLOOK}

In this paper we have begun the investigation of a class of spin systems
based on an arbitrary Lie group. The purpose of this study is to develop
a better understanding of the role of symmetry in the behaviour of spin
systems
in general. We believe that it may eventually be possible to generalize
some of the rigorous work that has traditionally been concentrated on the
$SU(2)$ systems, particularly that which concerns the nature of the
ground state: the question of long range order and its dependence on
parameters such as spin, coupling strength and the dimensionality of
space. At this point we have not addressed such questions. The work
described in this paper is of a preliminary nature, being concerned with
kinematics and only the simplest approximations: the correspondence
theory limit and the naive long wavelength limit. Even so, we believe
that the results have some interest.

Firstly, we have shown that the classical theory which emerges in the
correspondence theory limit is associated with a flag manifold or one
of its submanifolds. These manifolds are not, in general, symmetric
spaces although they are homogeneous and can be analyzed by group
theoretic methods. At least the preliminary features of the
semi--classical theory can be discerned in that it is possible to
identify dynamical variables and formulate a canonical quantization
programme. The problem of factor ordering, however, has been only
partially resolved. To express interesting Hamiltonians as well
defined self--adjoint and group invariant operators it will probably be
necessary to obtain more general versions of the Holstein--Primakoff
realization than we have found. Our version, discussed in Sec.6, is able
to realize only a limited  class of representations, those for which
the weights are not degenerate. These representations are
characterized by a single quantum number, $n_1-n_2$ in the case of
$SU(N+1)$. One of our motivations for going beyond $SU(2)$ was to have
a richer variety of quantum numbers on which interesting physical
quantities might depend. This remains to be achieved.

Secondly, we have shown that the long wavelength properties of the
classical theory are described by generalized non--relativistic
$\sigma$--models in which the fields take their values on a coset space
$G/H$ in which $H$ includes the maximal torus. The Lagrangian is of
first order in the time derivative but, depending on the nature of the
ground state ordering, it may be equivalent to a second order theory.
Such second order systems resemble the relativistic $\sigma$--models
except that the space and time derivatives are coupled to independent
tensors, $k_{\mu\nu}$ and $g_{\mu\nu}$, respectively. The flag
manifold generally admits more than one invariant tensor of rank 2.
This is to be contrasted with the homogeneous symmetric spaces which
admit only one: $\sigma$--models on symmetric spaces such as
$\bbc P^N$ are inevitably relativistic.

A question that calls for investigation is the long wavelength
behaviour of the quantized theory. At present it is not clear whether
the factor ordering ambiguities are significant in the continuum limit.
If they are not, then it should be possible to set up renormalization
group equations and search for fixed points. In particular, it would be
interesting, in the case of second order $\sigma$--models
to see whether the tensors $k_{\mu\nu}$ and
$g_{\mu\nu}$ evolve towards the relativistic form, $k_{\mu\nu}=
c^2g_{\mu\nu}$, at the fixed point. This idea is pursued in a separate
paper.

Finally, for generalized spin systems in one space dimension we have
shown that the Haldane topological term emerges in the continuum limit of
the antiferromagnetic chain, As in the $SU(2)$ case it appears that
representations for which the contravariant weight components,
$\Lambda^j$, are odd integers are distinguished. They contribute an
alternating sign to the path integrals.
\vfill\eject

\centerline{APPENDIX\qquad THE REALIZATION OF $SU(2)$}

A generalization of the Holstein--Primakoff realization can be
obtained by applying the method discussed in Sec.2. The structures here
are sufficiently simple that all formulae can be exhibited in detail
and so provide a useful illustration of the method. In particular, we
shall be able to see how the Holstein--Primakoff formulae emerge as
a special case.

Consider the $(2s+1)$--dimensional representation of $SU(2)$ defined
by
$$\eqalignno{
  J_3\ \vert m>\ &=\vert m>\ m\cr
  J_+\ \vert m>\ &=\vert m+1>\
  \sqrt{(s-m)(s+m+1)}\cr
  J_-\ \vert m>\ &= \vert m-1>\ \sqrt{(s+m)(s-m+1)}    &(A.1)\cr}$$
where $m=-s, -s+1,\dots ,s$ and $<m\vert m'>\ =\delta_{mm'}$. The
coherent states are defined by
$$\eqalignno{
  \vert\theta\varphi >\ &= L_{\theta\varphi}\ \vert\lambda >\cr
  &= e^{-i\varphi J_3}\ e^{-\theta J_2}\ e^{i\varphi J_3}\
  \vert\lambda >
&(A.2)\cr}$$
where $0\leq\theta <\pi$, $0\leq\varphi\leq 2\pi$ and $\vert\lambda >$
is one of the eigenstates of $J_3$. The expectation value of $\uJ$ is
a 3--vector of length $\vert\lambda\vert$,
$$\eqalignno{
  <\theta\varphi\vert J_\pm\vert\theta\varphi >\ &= \lambda\sin\theta
  \ e^{\pm i\varphi}\cr
  <\theta\varphi\vert J_3\vert\theta\varphi >\ &= \lambda\cos\theta\ .

   &(A.3)\cr}$$

Applying the methods of Sec.2, computing the infinitesimal everlap
of coherent states and expressing the transition amplitudes in path
integral notation one arrives at the classical Lagrangian
$$L=\barh\lambda (\cos\theta -1)\partial_t\varphi
  -H(\theta ,\varphi )
\eqno(A.4)$$
where $(\cos\theta -1)d\varphi$ is the spin connection on the 2--sphere
in polar coordinates. The canonical variables are
$$\varphi\quad {\rm and}\quad \pi =\barh\lambda (\cos\theta -1)\ .$$
In the quantized theory they are represented by operators subject to
the commutation rule
$$[\pi ,\varphi ]={\barh\over i}$$
or, more precisely, since states should be represented by periodic
functions of $\varphi$,
$$e^{i\varphi}\ \cos\theta\ e^{-i\varphi} =\cos\theta -{1\over\lambda}\ .

\eqno(A.5)$$
In the basis which diagonalizes $\varphi$ the eigenstates of $J_3$ are
represented by
$$<\varphi \vert m>\ = e^{im\varphi}\ .                     \eqno(A.6)$$
The canonical momemtum can be represented by the differential operator
$$\pi =-\barh (i\partial_\varphi +\lambda )                  \eqno(A.7)$$
which gives $\lambda\cos\theta =-i\partial_\varphi$. With this choice
one finds, using (A.1) and (A.6),
$$\eqalignno{
  J_3 &=\lambda\cos\theta\cr
  J_+ &= e^{i\varphi}\ \sqrt{(s-\lambda\cos\theta )(s+\lambda
  \cos\theta +1)}\cr
  &=\sqrt{(s-\lambda\cos\theta +1)(s+\lambda\cos\theta)}\
  e^{-i\varphi}\ .
&(A.8)\cr}$$
It is straightforward to verify that the operators (A.8) satisfy the
commutation rules of $SU(2)$ as well as the constraint
$${1\over 2} (J_+J_-+J_-J_+)+J^2_3=s(s+1)\ .               \eqno(A.9)$$

It is clear form (A.5) that the operators $\varphi$ and $\theta$
become commutative in the limit $\vert\lambda\vert\to\infty$.
Also the large quantum number behaviour of the operators (A.8)
approximates the classical formulae (A.3). For example, under the
assumptions
$$s^2-\lambda^2\mc O(\vert\lambda\vert )\quad {\rm and}\quad
  \vert\lambda\vert\sin^2\theta\gg 1                         \eqno(A.10)$$
as $s$ and $\vert\lambda\vert$ become large, one finds the
expansion
$$J_+\simeq\lambda\ e^{i\varphi}\sin\theta\left[ 1+
  {s(s+1)-\lambda^2-\lambda\cos\theta\over 2\lambda^2
  \sin^2\theta} -{1\over 8}\left(
  {s(s+1)-\lambda^2-\lambda\cos\theta\over
  \lambda^2\sin^2\theta}\right)^2+\dots\right]\ .             \eqno(A.11)$$

To recover the usual form of the Holstein--Primakoff realization define
the operator
$$\eqalignno{
  \phi &= e^{i\varphi}\ \sqrt{{1-\cos\theta\over 2}}\cr
  &=\sqrt{{1-{1\over\lambda}-\cos\theta\over 2}}\
  e^{i\varphi}
  &(A.12)\cr}$$
and its hermitian conjugate, $\phi^+$. The commutation rule becomes
$$[\phi ,\phi^+]=-{1\over 2\lambda}
\eqno(A.13)$$
which is the $SU(2)$ version of (6.10). The definition (A.12) can be
inverted to give
$$\cos\theta =1-2\phi^+\phi\quad {\rm and}\quad e^{-i\varphi}=
  \phi\ {1\over\sqrt{\phi^+\phi}} ={1\over\sqrt{\phi\phi^+}}\
  \phi\ .
         \eqno(A.14)$$
Hence, substituting into (A.8),
$$\eqalignno{
  J_3 &= \lambda -2\lambda\phi^+\phi\cr
  J_+ &=\phi^+\ \sqrt{{(s+\lambda -2\lambda\phi\phi^+)
  (s-\lambda +2\lambda\phi^+\phi)\over\phi\phi^+}}\cr
  &= \sqrt{{(s+1-\lambda +2\lambda\phi^+\phi )
  (s+\lambda -2\lambda\phi^+\phi)\over\phi^+\phi}}\
  \phi^+\cr
  J_- &= J^+_+\ .
  &(A.15)\cr}$$
The singularity at $\phi =0$ in these formulae is eliminated by
choosing $s=\pm\ \lambda$. For example, with $\lambda =-s$, (A.15)
reduces to the familiar form
$$\eqalignno{
  J_3 &=-s+2s\phi^+\phi\cr
  J_+ &=2s\ \phi^+\ \sqrt{1-\phi^+\phi}\cr
  &=2s\ \sqrt{1+{1\over 2s} -\phi^+\phi}\ \phi^+            &(A.16)\cr}$$
with $[\phi ,\phi^+]=1/2s$. This is the $SU(2)$ version of the
realization (6.12). On the basis of this simple example we
are encouraged to believe
that it may be possible to generalize the operators (6.12) and (6.14)
to yield the representation $(n_1,n_2)$ of $SU(N+1)$.
\vfill\eject

\centerline{REFERENCES}
\bigskip

\item{1)}
For reviews see, for example,\hfil\break
I. Affleck, in {\it Strings, Fields and Critical Phenomena}, Les Houches
Summer School, 1988, Session XLIX, edited by E. Brezin and
J. Zinn--Justin (North Holland, 1990);\hfil\break
E. Fradkin, {\it Field Theories of Condensed Matter Systems}
(Addison--Wesley, Redwood City, 1991);\hfil\break
E. Manousakis, Rev. Mod. Phys. {\bf 63}, 1 (1991).

\item{2)}
I. Affleck, Nucl. Phys. {\bf B257}, 397 (1985);\hfil\break
I. Affleck, Nucl. Phys. {\bf B265}, 409 (1986).

\item{3)}
H.A. Bethe, Z. Phys. {\bf 71}, 205 (1931);\hfil\break
R.L. Orbach, Phys. Rev. {\bf 112}, 309 (1958).

\item{4)}
C. Kittel, {\it Quantum Theory of Solids} (Wiley, New York, 1963).

\item{5)}
F.D.M. Haldane, Phys. Lett. {\bf 93A}, 464 (1983); Phys. Rev. Lett.
{\bf 50}, 1153 (1983);\hfil\break
See also Ref.2.

\item{6)}
See, for example, the discussion in Sec.1 of\hfil\break
S. Chakrabarty, B.I. Halperin and D.R. Nelson, Phys. Rev. {\bf B39},
2344 (1989).

\item{7)}
See, for example, the review by Manousaki, Ref.1.

\item{8)}
F.D.M. Haldane, Phys. Rev. Lett. {\bf 57}, 1488 (1986);\hfil\break
See also A. Perelomov, {\it Generalized Coherent States and their
Applications} (Springer, 1986).

\item{9)}
F.D.M. Haldane, J. Appl. Phys. {\bf 57}, 3359 (1985).

\item{10)}
E. Fradkin and M. Stone, Phys. Rev. {\bf B38}, 7215 (1988);\hfil\break
F.D.M. Haldane, Phys. Rev. Lett. {\bf 61}, 1029 (1988);\hfil\break
X.G. Wen and A. Zee, Phys. Rev. Lett. {\bf 61}, 1025 (1988);\hfil\break
D. Dombre and N. Read, Phys. Rev. {\bf B38}, 7181 (1988).

\item{11)}
T. Holstein and H. Primakoff, Phys. Rev. {\bf 58}, 1098 (1940).

\item{12)}
A. Belavin and A. Polyakov, JETP Lett. {\bf 22} (1975) 245;\hfil\break
A. D'Adda, P. di Vecchia and M. L\"uscher, Nucl. PHys. {\bf 146B}, 63
(1978);\hfil\break
M. L\"uscher, Nucl. Phys. {\bf 135B}, 1 (1978);\hfil\break
H. Eichenherr, Nucl. PHys. {\bf 146B}, 250 (1978): {\bf 155B}, 544 (1979).

\item{13)}
E.M. Lifshitz and L.P. Pitaevskii, {\it Statistical Physics, Part 2}
(Pergamon Press, Oxford, 1980).
\vfill\eject
\bye